%% file: main.tex
\documentclass[sigplan,10pt,nonacm]{acmart}
\renewcommand\footnotetextcopyrightpermission[1]{}
\usepackage[]{hyperref}
\input{preamble.tex}

\begin{document}

\title{\system: A Fine-Grained GPU Memory Sharing System for LLM Serving with Opportunistic Fine-Tuning}

\author{Jaehoon Yang}
\affiliation{\institution{Seoul National University}\city{Seoul}\country{Republic of Korea}}
\email{jaehoon.yang@snu.ac.kr}
\author{Yongbeom Kim}
\affiliation{\institution{Seoul National University}\city{Seoul}\country{Republic of Korea}}
\email{kybs0627@snu.ac.kr}
\author{Hojoon Kim}
\affiliation{\institution{Seoul National University}\city{Seoul}\country{Republic of Korea}}
\email{hojoon.kim@snu.ac.kr}
\author{Seung Yul Lee}
\affiliation{\institution{Seoul National University}\city{Seoul}\country{Republic of Korea}}
\email{triomphant1@snu.ac.kr}
\author{Jae W. Lee}
\affiliation{\institution{Seoul National University}\city{Seoul}\country{Republic of Korea}}
\email{jaewlee@snu.ac.kr}
\renewcommand{\shortauthors}{Yang et al.}

\begin{abstract}
Large language model (LLM) serving scales its replica count with the request load, yet GPU memory still stands idle inside the replicas. Adding a replica takes minutes, while the memory that a replica needs changes within seconds. Even instant autoscaling could not return this idle memory, because the smallest unit that it can remove is a whole replica. Colocating parameter-efficient fine-tuning (PEFT) with inference can use this memory, but inference must be able to reclaim it within seconds, before requests that wait for memory exceed their latency service-level objective (SLO). Existing colocation systems either keep the tuning memory resident or let inference reclaim it at the coarse granularity of a whole training sample. Each such reclamation also discards the running tuning step. To address these limitations, we present \system, a fine-grained memory sharing system that lets inference reclaim the memory of individual activations that a running tuning step has saved for its backward pass. The step continues, and its backward pass recomputes those activations. Inference reclaims only memory that no in-flight GPU work can still access, even under CPU--GPU asynchrony and tensor parallelism. On \rHeadEvalScope, \system keeps inference SLO attainment at or above \rHeadSloMolt and completes \rHeadTputGain the tuning work of discard-based memory sharing.
\end{abstract}

\maketitle

\typeout{TW=\the\textwidth CW=\the\columnwidth}

\input{contents/1-intro}

\input{contents/2-background}
\input{contents/3-motivation}

\input{contents/4a-overview}
\input{contents/4b-design}
\input{contents/5-evaluation}
\input{contents/6-related}
\input{contents/7-con}

\bibliographystyle{ACM-Reference-Format}
\bibliography{refs.bib}

\end{document}

%% file: preamble.tex
\usepackage{multirow}
\usepackage{xspace}
\usepackage{tikz}
\usepackage{array}
\usepackage{listings}
\usepackage{caption}
\usepackage{subcaption}
\newlength{\trimSiriusLimitation}
\newlength{\trimBurst}
\newlength{\trimFootprint}
\newlength{\trimActLevelMem}
\newlength{\trimHandoverMain}
\newlength{\trimHandoverContrast}
\newlength{\trimAsyncThrottling}
\newlength{\trimAsyncTp}
\newlength{\trimOverview}
\newlength{\trimMemutil}
\newlength{\trimCapsweep}
\newlength{\trimHoover}
\newlength{\skipBgTraces}
\usepackage{afterpage}
\AtBeginDocument{\clubpenalty=1000 \widowpenalty=1000 }
\newlength{\floatGapPageOne}
\newlength{\floatGapDefault}
\AtBeginDocument{\setlength{\floatGapDefault}{\textfloatsep}}
\newcommand{\pageOneFloatGap}{\setlength{\textfloatsep}{\floatGapPageOne}\afterpage{\global\textfloatsep=\floatGapDefault\relax}}
\newcommand{\flushList}[1]{\settowidth{\leftmargini}{#1}\addtolength{\leftmargini}{\labelsep}}
\usepackage[ruled,vlined,linesnumbered]{algorithm2e}
\usepackage{amsmath}
\usepackage{booktabs}
\usepackage{tabularx}
\usepackage{xcolor}
\usepackage[normalem]{ulem}     %

\input{framework/latex/markers.tex}       %
\input{framework/latex/review-marks.tex}  %

\definecolor{revpurple}{HTML}{CC79A7}

\input{macros.tex}                        %
\input{results_macros.tex}                %

\definecolor{mauvelous}{rgb}{0.94, 0.6, 0.67}
\lstdefinestyle{myStyle1}{
  belowcaptionskip=1\baselineskip,
  frame=tb,
  language=c++,
  aboveskip=0mm,
  belowskip=0mm,
  showstringspaces=false,
  columns=flexible,
  basicstyle={\fontsize{8.0pt}{8.0pt}\fontfamily{pcr}\selectfont},
  numbers=left,
  xleftmargin=2em,
  numberstyle={\color{gray}\texttt},
  keywordstyle=\color{black}\textbf,
  commentstyle=\color{mauvelous},
  stringstyle=\color{mauvelous},
  frame=none,
  breaklines=true,
  breakatwhitespace=true,
  tabsize=2,
  morekeywords={WHILE, IF, FOR, ELSE, IS, IN, TO, FROM},
  deletekeywords={and, sizeof},
}

%% file: framework/latex/markers.tex
\usepackage{xcolor}

\definecolor{phgreen}{HTML}{009E73}

%% file: framework/latex/review-marks.tex
\definecolor{revorange}{HTML}{D95F02}
\definecolor{revblue}{HTML}{0072B2}
\definecolor{revgreen}{HTML}{009E73}

%% file: macros.tex
\newcommand{\system}{MOLT\xspace}
\newcommand{\sirius}{SIRIUS\xspace}
\newcommand{\llmstation}{LLMStation\xspace}

\newcommand{\schemeA}{flexible activation checkpointing\xspace}
\newcommand{\schemeB}{continuous memory handover\xspace}
\newcommand{\schemeC}{asynchronous-aware slot protection\xspace}
\newcommand{\moltcont}{\system with continuous handover\xspace}
\newcommand{\moltstep}{\system with step-level handover\xspace}
\newcommand{\schemeATitle}{Flexible Activation Checkpointing\xspace}
\newcommand{\schemeCTitle}{Asynchronous-aware Slot Protection\xspace}

\newcommand{\sst}{slot state table\xspace}
\newcommand{\sstshort}{SST\xspace}
\newcommand{\save}{save-path\xspace}
\newcommand{\recomp}{recompute-path\xspace}
\newcommand{\reclaimable}{reclaimable pool\xspace}
\newcommand{\pinned}{pinned pool\xspace}

\newcommand{\mistralsmall}{Mistral-Small-24B-Instruct\xspace}
\newcommand{\llama}{Llama-3.1-70B-Instruct\xspace}

\newcommand{\gpuh}{H100\xspace}
\newcommand{\gpub}{B200\xspace}
\newcommand{\gpuhsxm}{H100 SXM\xspace}

%% file: results_macros.tex
\newif\ifvaluehl
\valuehlfalse
\newcommand{\vhl}[1]{\ifvaluehl\textcolor{red}{#1}\else#1\fi}

\newcommand{\rSetupLocLib}{\vhl{13.1K}\xspace}
\newcommand{\rSetupVllmPatchFiles}{\vhl{23}\xspace}
\newcommand{\rSetupVllmPatchLines}{\vhl{2.7K}\xspace}
\newcommand{\rSetupVllmModule}{\vhl{9.7K}\xspace}

\newcommand{\rSetupVllm}{\vhl{v0.11.0}\xspace}

\newcommand{\rSetupTorch}{\vhl{2.8.0}\xspace}

\newcommand{\rSetupLayersSmall}{\vhl{40}\xspace}
\newcommand{\rSetupLayersLlama}{\vhl{80}\xspace}
\newcommand{\rSetupTpSmall}{\vhl{1}\xspace}
\newcommand{\rSetupTpLlama}{\vhl{1}\xspace}
\newcommand{\rSetupTpLarge}{\vhl{2}\xspace}

\newcommand{\rSetupDeploySmall}{\vhl{1$\times$\gpuhsxm}}
\newcommand{\rSetupDeployLlama}{\vhl{1$\times$\gpub}}
\newcommand{\rSetupDeployLlamaSxm}{\vhl{4$\times$\gpuhsxm}}
\newcommand{\rSetupTpLlamaSxm}{\vhl{4}\xspace}
\newcommand{\rSetupDeploySmallTptwo}{\vhl{2$\times$\gpuhsxm}}
\newcommand{\rSetupTpSmallTptwo}{\vhl{2}\xspace}

\newcommand{\rSetupSeqSmall}{\vhl{768}\xspace}
\newcommand{\rSetupSeqSmallTptwo}{\vhl{2048}\xspace}
\newcommand{\rSetupSeqLlamaSxm}{\vhl{1024}\xspace}
\newcommand{\rSetupSeqLlama}{\vhl{512}\xspace}

\newcommand{\rSetupLoraRank}{\vhl{8}\xspace}

\newcommand{\rSetupFactorMin}{\vhl{2.0}\xspace}
\newcommand{\rSetupFactorMax}{\vhl{4.0}\xspace}

\newcommand{\rSetupSloMultStrict}{\vhl{3$\times$}\xspace}

\newcommand{\rMotiBatchOneFootprintSeq}{\vhl{1024 tokens}\xspace}
\newcommand{\rMotiBatchOneFootprintGiB}{\vhl{15.8\,GiB}\xspace}
\newcommand{\rMotiPersistentGiB}{\vhl{0.4\,GiB}\xspace}

\newcommand{\rMotiAdaptersTotal}{\vhl{256{,}295}\xspace}
\newcommand{\rMotiAdaptersSpan}{\vhl{December 2023 and September 2026}\xspace}

\newcommand{\rMotiAdaptersLlama}{\vhl{37{,}606}\xspace}

\newcommand{\rMotiDiurnalPeakMeanConv}{\vhl{1.7$\times$}\xspace}
\newcommand{\rMotiDiurnalPeakMeanCode}{\vhl{3.0$\times$}\xspace}
\newcommand{\rMotiDiurnalPeakMeanBurst}{\vhl{4.1$\times$}\xspace}

\newcommand{\rMotiRateIntervalSwing}{\vhl{1.25$\times$}\xspace}

\newcommand{\rMotiKvRangeGiB}{\vhl{3.9--14.8\,GiB}\xspace}
\newcommand{\rMotiKvSwingWindow}{\vhl{ten seconds}\xspace}
\newcommand{\rMotiKvWindowSeconds}{\vhl{227\,s}\xspace}
\newcommand{\rMotiKvRiseTen}{\vhl{3.2$\times$}\xspace}

\newcommand{\rMotiWeightsGiB}{\vhl{43.9\,GiB}\xspace}
\newcommand{\rMotiKvIdleGiB}{\vhl{20--31\,GiB}\xspace}

\newcommand{\rMotiKvIdleGiBTp}{\vhl{32--44\,GiB}\xspace}
\newcommand{\rMotiKvPeakTotalGiB}{\vhl{51\,GiB}\xspace}
\newcommand{\rMotiPoolTpOneGiB}{\vhl{25\,GiB}\xspace}
\newcommand{\rMotiPoolTpTwoGiB}{\vhl{87\,GiB}\xspace}

\newcommand{\rMotiAutoscaleTarget}{\vhl{60\%}\xspace}
\newcommand{\rMotiAutoscaleHeadroom}{\vhl{40\%}\xspace}

\newcommand{\rEteSxmCoarsePeft}{\vhl{212.8}\xspace}
\newcommand{\rEteSxmCoarseSlo}{\vhl{69.0}\xspace}
\newcommand{\rEteSxmCoarseTpotMed}{\vhl{48.3}\xspace}
\newcommand{\rEteSxmCoarseTpotPninetyfive}{\vhl{62.7}\xspace}
\newcommand{\rEteSxmCoarseTpotPseventyfive}{\vhl{52.8}\xspace}
\newcommand{\rEteSxmCoarseTtftMed}{\vhl{585.2}\xspace}
\newcommand{\rEteSxmCoarseTtftPninetyfive}{\vhl{3112.9}\xspace}
\newcommand{\rEteSxmCoarseTtftPseventyfive}{\vhl{1208.0}\xspace}
\newcommand{\rEteSxmMoltPeft}{\vhl{242.6}\xspace}
\newcommand{\rEteSxmMoltSlo}{\vhl{99.7}\xspace}
\newcommand{\rEteSxmMoltTpotMed}{\vhl{47.0}\xspace}
\newcommand{\rEteSxmMoltTpotPninetyfive}{\vhl{56.6}\xspace}
\newcommand{\rEteSxmMoltTpotPseventyfive}{\vhl{51.1}\xspace}
\newcommand{\rEteSxmMoltTtftMed}{\vhl{368.5}\xspace}
\newcommand{\rEteSxmMoltTtftPninetyfive}{\vhl{619.7}\xspace}
\newcommand{\rEteSxmMoltTtftPseventyfive}{\vhl{449.2}\xspace}

\newcommand{\rEteSxmLmsPeft}{\vhl{285.1}\xspace}
\newcommand{\rEteSxmLmsSlo}{\vhl{0.0}\xspace}
\newcommand{\rEteSxmLmsTpotMed}{\vhl{42.8}\xspace}
\newcommand{\rEteSxmLmsTpotPninetyfive}{\vhl{49.0}\xspace}
\newcommand{\rEteSxmLmsTpotPseventyfive}{\vhl{44.6}\xspace}
\newcommand{\rEteSxmLmsTtftMed}{\vhl{18671.4}\xspace}
\newcommand{\rEteSxmLmsTtftPninetyfive}{\vhl{26935.0}\xspace}
\newcommand{\rEteSxmLmsTtftPseventyfive}{\vhl{24154.1}\xspace}
\newcommand{\rEteSxmSiriusPeft}{\vhl{127.7}\xspace}
\newcommand{\rEteSxmSiriusSlo}{\vhl{99.2}\xspace}
\newcommand{\rEteSxmSiriusTpotMed}{\vhl{39.7}\xspace}
\newcommand{\rEteSxmSiriusTpotPninetyfive}{\vhl{55.4}\xspace}
\newcommand{\rEteSxmSiriusTpotPseventyfive}{\vhl{43.1}\xspace}
\newcommand{\rEteSxmSiriusTtftMed}{\vhl{384.9}\xspace}
\newcommand{\rEteSxmSiriusTtftPninetyfive}{\vhl{723.6}\xspace}
\newcommand{\rEteSxmSiriusTtftPseventyfive}{\vhl{465.7}\xspace}

\newcommand{\rEteSxmTptwoCoarsePeft}{\vhl{509.2}\xspace}
\newcommand{\rEteSxmTptwoCoarseSlo}{\vhl{77.1}\xspace}
\newcommand{\rEteSxmTptwoCoarseTpotMed}{\vhl{62.6}\xspace}
\newcommand{\rEteSxmTptwoCoarseTpotPninetyfive}{\vhl{85.8}\xspace}
\newcommand{\rEteSxmTptwoCoarseTpotPseventyfive}{\vhl{71.6}\xspace}
\newcommand{\rEteSxmTptwoCoarseTtftMed}{\vhl{323.5}\xspace}
\newcommand{\rEteSxmTptwoCoarseTtftPninetyfive}{\vhl{1544.0}\xspace}
\newcommand{\rEteSxmTptwoCoarseTtftPseventyfive}{\vhl{606.4}\xspace}
\newcommand{\rEteSxmTptwoMoltPeft}{\vhl{557.7}\xspace}
\newcommand{\rEteSxmTptwoMoltSlo}{\vhl{100.0}\xspace}
\newcommand{\rEteSxmTptwoMoltTpotMed}{\vhl{68.4}\xspace}
\newcommand{\rEteSxmTptwoMoltTpotPninetyfive}{\vhl{84.2}\xspace}
\newcommand{\rEteSxmTptwoMoltTpotPseventyfive}{\vhl{74.7}\xspace}
\newcommand{\rEteSxmTptwoMoltTtftMed}{\vhl{284.9}\xspace}
\newcommand{\rEteSxmTptwoMoltTtftPninetyfive}{\vhl{468.0}\xspace}
\newcommand{\rEteSxmTptwoMoltTtftPseventyfive}{\vhl{342.7}\xspace}

\newcommand{\rEteSxmTptwoLmsPeft}{\vhl{606.1}\xspace}
\newcommand{\rEteSxmTptwoLmsSlo}{\vhl{5.0}\xspace}
\newcommand{\rEteSxmTptwoLmsTpotMed}{\vhl{60.6}\xspace}
\newcommand{\rEteSxmTptwoLmsTpotPninetyfive}{\vhl{68.8}\xspace}
\newcommand{\rEteSxmTptwoLmsTpotPseventyfive}{\vhl{64.1}\xspace}
\newcommand{\rEteSxmTptwoLmsTtftMed}{\vhl{6746.8}\xspace}
\newcommand{\rEteSxmTptwoLmsTtftPninetyfive}{\vhl{10359.7}\xspace}
\newcommand{\rEteSxmTptwoLmsTtftPseventyfive}{\vhl{8188.0}\xspace}
\newcommand{\rEteSxmTptwoSiriusPeft}{\vhl{242.5}\xspace}
\newcommand{\rEteSxmTptwoSiriusSlo}{\vhl{95.5}\xspace}
\newcommand{\rEteSxmTptwoSiriusTpotMed}{\vhl{49.3}\xspace}
\newcommand{\rEteSxmTptwoSiriusTpotPninetyfive}{\vhl{83.3}\xspace}

\newcommand{\rEteSxmTptwoSiriusTpotPseventyfive}{\vhl{58.4}\xspace}
\newcommand{\rEteSxmTptwoSiriusTtftMed}{\vhl{278.4}\xspace}
\newcommand{\rEteSxmTptwoSiriusTtftPninetyfive}{\vhl{674.6}\xspace}
\newcommand{\rEteSxmTptwoSiriusTtftPseventyfive}{\vhl{385.1}\xspace}

\newcommand{\rEteLlamaBMoltSlo}{\vhl{100.0}\xspace}
\newcommand{\rEteLlamaBMoltPeft}{\vhl{90.3}\xspace}
\newcommand{\rEteLlamaBMoltTtftMed}{\vhl{592.4}\xspace}
\newcommand{\rEteLlamaBMoltTtftPseventyfive}{\vhl{735.5}\xspace}
\newcommand{\rEteLlamaBMoltTtftPninetyfive}{\vhl{985.8}\xspace}
\newcommand{\rEteLlamaBMoltTpotMed}{\vhl{63.8}\xspace}
\newcommand{\rEteLlamaBMoltTpotPseventyfive}{\vhl{67.1}\xspace}
\newcommand{\rEteLlamaBMoltTpotPninetyfive}{\vhl{73.9}\xspace}
\newcommand{\rEteLlamaBCoarseSlo}{\vhl{74.3}\xspace}
\newcommand{\rEteLlamaBCoarsePeft}{\vhl{69.6}\xspace}
\newcommand{\rEteLlamaBCoarseTtftMed}{\vhl{825.2}\xspace}
\newcommand{\rEteLlamaBCoarseTtftPseventyfive}{\vhl{1822.1}\xspace}
\newcommand{\rEteLlamaBCoarseTtftPninetyfive}{\vhl{6219.3}\xspace}
\newcommand{\rEteLlamaBCoarseTpotMed}{\vhl{61.4}\xspace}
\newcommand{\rEteLlamaBCoarseTpotPseventyfive}{\vhl{67.7}\xspace}
\newcommand{\rEteLlamaBCoarseTpotPninetyfive}{\vhl{82.1}\xspace}
\newcommand{\rEteLlamaBLmsSlo}{\vhl{0.0}\xspace}
\newcommand{\rEteLlamaBLmsPeft}{\vhl{76.1}\xspace}
\newcommand{\rEteLlamaBLmsTtftMed}{\vhl{40556.2}\xspace}
\newcommand{\rEteLlamaBLmsTtftPseventyfive}{\vhl{52458.2}\xspace}
\newcommand{\rEteLlamaBLmsTtftPninetyfive}{\vhl{59190.3}\xspace}
\newcommand{\rEteLlamaBLmsTpotMed}{\vhl{56.1}\xspace}
\newcommand{\rEteLlamaBLmsTpotPseventyfive}{\vhl{58.9}\xspace}
\newcommand{\rEteLlamaBLmsTpotPninetyfive}{\vhl{63.4}\xspace}
\newcommand{\rEteLlamaBSiriusSlo}{\vhl{99.7}\xspace}
\newcommand{\rEteLlamaBSiriusPeft}{\vhl{42.6}\xspace}
\newcommand{\rEteLlamaBSiriusTtftMed}{\vhl{595.5}\xspace}
\newcommand{\rEteLlamaBSiriusTtftPseventyfive}{\vhl{763.6}\xspace}
\newcommand{\rEteLlamaBSiriusTtftPninetyfive}{\vhl{976.9}\xspace}
\newcommand{\rEteLlamaBSiriusTpotMed}{\vhl{54.5}\xspace}
\newcommand{\rEteLlamaBSiriusTpotPseventyfive}{\vhl{58.2}\xspace}
\newcommand{\rEteLlamaBSiriusTpotPninetyfive}{\vhl{66.0}\xspace}

\newcommand{\rEteLlamaSxmMoltSlo}{\vhl{99.7}\xspace}
\newcommand{\rEteLlamaSxmMoltPeft}{\vhl{64.5}\xspace}
\newcommand{\rEteLlamaSxmMoltTtftMed}{\vhl{466.4}\xspace}
\newcommand{\rEteLlamaSxmMoltTtftPseventyfive}{\vhl{569.3}\xspace}
\newcommand{\rEteLlamaSxmMoltTtftPninetyfive}{\vhl{811.4}\xspace}
\newcommand{\rEteLlamaSxmMoltTpotMed}{\vhl{107.6}\xspace}
\newcommand{\rEteLlamaSxmMoltTpotPseventyfive}{\vhl{118.3}\xspace}
\newcommand{\rEteLlamaSxmMoltTpotPninetyfive}{\vhl{136.8}\xspace}
\newcommand{\rEteLlamaSxmCoarseSlo}{\vhl{74.3}\xspace}
\newcommand{\rEteLlamaSxmCoarsePeft}{\vhl{64.5}\xspace}
\newcommand{\rEteLlamaSxmCoarseTtftMed}{\vhl{558.9}\xspace}
\newcommand{\rEteLlamaSxmCoarseTtftPseventyfive}{\vhl{1317.4}\xspace}
\newcommand{\rEteLlamaSxmCoarseTtftPninetyfive}{\vhl{3543.5}\xspace}
\newcommand{\rEteLlamaSxmCoarseTpotMed}{\vhl{106.9}\xspace}
\newcommand{\rEteLlamaSxmCoarseTpotPseventyfive}{\vhl{118.2}\xspace}
\newcommand{\rEteLlamaSxmCoarseTpotPninetyfive}{\vhl{142.5}\xspace}
\newcommand{\rEteLlamaSxmLmsSlo}{\vhl{2.8}\xspace}
\newcommand{\rEteLlamaSxmLmsPeft}{\vhl{83.8}\xspace}
\newcommand{\rEteLlamaSxmLmsTtftMed}{\vhl{14030.5}\xspace}
\newcommand{\rEteLlamaSxmLmsTtftPseventyfive}{\vhl{20267.1}\xspace}
\newcommand{\rEteLlamaSxmLmsTtftPninetyfive}{\vhl{27178.1}\xspace}
\newcommand{\rEteLlamaSxmLmsTpotMed}{\vhl{101.1}\xspace}
\newcommand{\rEteLlamaSxmLmsTpotPseventyfive}{\vhl{106.0}\xspace}
\newcommand{\rEteLlamaSxmLmsTpotPninetyfive}{\vhl{113.8}\xspace}
\newcommand{\rEteLlamaSxmSiriusSlo}{\vhl{100.0}\xspace}
\newcommand{\rEteLlamaSxmSiriusPeft}{\vhl{19.3}\xspace}
\newcommand{\rEteLlamaSxmSiriusTtftMed}{\vhl{448.2}\xspace}
\newcommand{\rEteLlamaSxmSiriusTtftPseventyfive}{\vhl{513.2}\xspace}
\newcommand{\rEteLlamaSxmSiriusTtftPninetyfive}{\vhl{705.3}\xspace}
\newcommand{\rEteLlamaSxmSiriusTpotMed}{\vhl{79.4}\xspace}
\newcommand{\rEteLlamaSxmSiriusTpotPseventyfive}{\vhl{87.0}\xspace}
\newcommand{\rEteLlamaSxmSiriusTpotPninetyfive}{\vhl{105.1}\xspace}

\newcommand{\rEteMoltOverSiriusRange}{\vhl{1.9--3.3$\times$}\xspace}

\newcommand{\rAblFactorCount}{\vhl{five}\xspace}

\newcommand{\rAblBestFactor}{\vhl{3.0}\xspace}
\newcommand{\rAblWorstSlo}{\vhl{43.1}\xspace}

\newcommand{\rAblHandoversFine}{\vhl{194}\xspace}
\newcommand{\rAblHandoversCoarse}{\vhl{24}\xspace}

\newcommand{\rMemPlotSpan}{\vhl{80\,s}\xspace}
\newcommand{\rMemSiriusUnusedPct}{\vhl{87\%}\xspace}
\newcommand{\rMemMoltBackedMeanPct}{\vhl{86\%}\xspace}

\newcommand{\rMemCoarseBackedMeanPct}{\vhl{81\%}\xspace}

\newcommand{\rMemQueuedPctCoarse}{\vhl{47\%}\xspace}

\newcommand{\rMemQueuedPctCoarseOthers}{\vhl{54--64\%}\xspace}
\newcommand{\rMemQueuedPctMoltMax}{\vhl{1.4\%}\xspace}
\newcommand{\rMemQueuedPctCoarseRange}{\vhl{27--47\%}\xspace}
\newcommand{\rSetupPollInterval}{\vhl{0.5\,s}\xspace}

\newcommand{\rMemMoltTrackLatency}{\vhl{1.6\,ms}\xspace}

\newcommand{\rMemWastedFracSirius}{\vhl{16.7\%}\xspace}

\newcommand{\rCapBackwardFloor}{\vhl{five}\xspace}

\newcommand{\rCapBandCaps}{\vhl{32, 64, 128 and 32}\xspace}

\newcommand{\rCapClaimedAtBandPct}{\vhl{12.5, 12.5, 7.5 and 6.3\%}\xspace}
\newcommand{\rCapClaimedNoCapPct}{\vhl{45, 65, 100 and 55\%}\xspace}
\newcommand{\rCapStepTightRange}{\vhl{13.1--24.6\%}\xspace}

\newcommand{\rCapTightest}{\vhl{4}\xspace}

\newcommand{\rCapTolerancePct}{\vhl{5\%}\xspace}

\newcommand{\rHoRepsMin}{\vhl{19}\xspace}

\newcommand{\rHoSizeDrawnList}{\vhl{256, 128, 128 and 512\,MiB}\xspace}

\newcommand{\rHoLargestShareList}{\vhl{1.0, 1.3, 0.7 and 0.8\%}\xspace}

\newcommand{\rHoFixedTpOne}{\vhl{0.64--0.87\,ms}\xspace}
\newcommand{\rHoFixedTpMulti}{\vhl{1.09--1.47\,ms}\xspace}
\newcommand{\rHoAllocFlatTpOne}{\vhl{0.44--0.72\,ms}\xspace}
\newcommand{\rHoAllocDoublingTpOne}{\vhl{1.5--2.0}\xspace}
\newcommand{\rHoAllocKneeSize}{\vhl{64\,MiB}\xspace}
\newcommand{\rHoPaceSizeLlamaSxm}{\vhl{64\,MiB}\xspace}
\newcommand{\rHoWorkLlamaSxm}{\vhl{0.38--0.53\,ms}\xspace}
\newcommand{\rHoPlanTpOne}{\vhl{0.05--0.12\,ms}\xspace}
\newcommand{\rHoPlanTpTwo}{\vhl{0.22--0.32\,ms}\xspace}
\newcommand{\rHoPlanStepPct}{\vhl{0.07\%}\xspace}

\newcommand{\rHeadTputGain}{\vhl{1.9--3.3$\times$}\xspace}  %
\newcommand{\rHeadHoOverhead}{\vhl{1.3\%}\xspace}                  %

\newcommand{\rHeadEvalScope}{\vhl{four model deployments (24B--70B) across
  \gpuhsxm and \gpub GPUs under trace-driven
  workloads}\xspace}
\newcommand{\rHeadSloMolt}{\vhl{99.7\%}\xspace}
\newcommand{\rHeadSloLms}{\vhl{5.0\%}\xspace}

%% file: contents/1-intro.tex
\section{Introduction}
\label{sec:intro}

\begin{figure}[t]
  \centering
  \includegraphics[page=1, width=\columnwidth, trim=0 {\trimSiriusLimitation} 0 0]{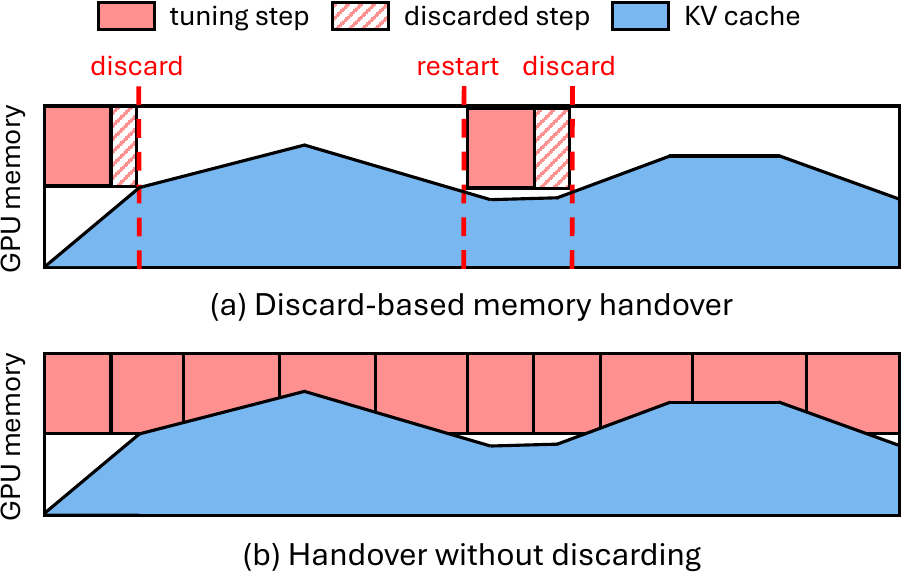}
  \caption{Tuning work that discard-based handover discards and handover without discarding keeps.}
  \label{fig:sirius-limitation}
\end{figure}
\pageOneFloatGap

Request load in large language model (LLM) serving fluctuates heavily. Provisioning enough GPUs to meet the latency service-level objective (SLO) at peak load leaves them underutilized the rest of the time, so production systems autoscale. They add and remove replicas, each an independent serving instance that holds one copy of the model on one or more GPUs, as the load changes~\cite{gkeautoscale,vertexautoscale}. Autoscaling follows the slow part of this fluctuation well, the daily cycle in which user-facing services peak during working hours and fall at night~\cite{burstgpt,dynamollm}, and removes the capacity that a lower load leaves idle~\cite{sageserve}. The fast part of the fluctuation, however, outpaces autoscaling, since the key-value (KV) cache of a replica, which dominates its dynamic memory use, can grow \rMotiKvRiseTen within \rMotiKvSwingWindow while bringing up a replica and loading the model takes up to several minutes~\cite{dynamollm}. Each replica must therefore keep headroom to absorb the load that arrives before new replicas come up. One production autoscaler targets \rMotiAutoscaleTarget GPU utilization by default, leaving each GPU idle \rMotiAutoscaleHeadroom of the time~\cite{vertexautoscale}.
A deployment thus runs at an \emph{autoscaling floor}, the fewest replicas that keep this headroom, where GPU compute and memory stay idle even with autoscaling. This memory cannot be returned to the provider, because a replica is withdrawn only as a whole.

Colocating another workload with inference can put this idle capacity to use. Parameter-efficient fine-tuning (PEFT) is a natural candidate. It is the standard way to adapt a deployed model~\cite{peftsurvey}, and demand keeps growing, with Hugging Face hosting \rMotiAdaptersTotal PEFT adapters published at a rising rate since 2024~\cite{llmstation}. 
It also fits colocation well. PEFT freezes the pretrained weights and trains only a small adapter~\cite{lora}, so when applied to the model that inference is serving it reuses the weights resident on the GPU.
Its progress is measured in throughput rather than latency, so it can yield memory to inference during a burst and take it back afterwards.

Existing colocation systems dynamically share either the compute or the memory of the GPU.
\llmstation~\cite{llmstation} shares compute by interleaving tuning operations with inference iterations, but its tuning memory stays resident.
Memory, however, is what determines whether inference meets its SLO during a burst. 
Inference admits a request only when the KV cache has room for it~\cite{vllm,orca}, so a shortage of memory delays admission and lengthens the time to first token (TTFT). Likewise, a running request whose KV cache cannot grow for lack of memory stalls until memory frees up~\cite{vllm,llumnix}.
\sirius~\cite{sirius} focuses on sharing memory, treating the training batch as elastic. When inference needs memory, it aborts the running step, restarts it with a smaller batch, and hands the freed memory to inference in a \emph{handover} fast enough for the SLO. The memory boundary between the two workloads thus moves at runtime.

\sirius uses the idle memory that autoscaling removes, the memory a fixed number of GPUs leaves free as the load changes over minutes and hours. At the autoscaling floor, however, resizing the training batch moves the boundary in increments too coarse and at a cost too high. First, the smallest increment is one training sample, whose memory is large relative to the idle memory. A one-sample \mistralsmall tuning step at \rMotiBatchOneFootprintSeq holds \rMotiBatchOneFootprintGiB beside the frozen weights, against \rMotiKvIdleGiB idle on one \gpuh. Tuning halts once the idle memory cannot hold one sample. Second, each handover discards the running tuning step and restarts it with a smaller batch, losing the work already done (Figure~\ref{fig:sirius-limitation}(a)). At the floor, this cost recurs whenever the KV cache outgrows the free memory, which happens within seconds.

Avoiding both limitations requires a unit of memory finer than a training sample, one that inference can reclaim without discarding the running step. We observe that the saved activation is such a unit. A saved activation is a tensor that the forward pass keeps for the backward pass to compute gradients. Saved activations hold most of a PEFT step's memory~\cite{meft} and can be recovered by recomputation~\cite{chen2016,checkmate,dtr}. When inference reclaims the memory of some of them, the step continues, and its backward pass recomputes them (Figure~\ref{fig:sirius-limitation}(b)). Recomputation itself is well studied, but activation checkpointing fixes its memory budget before tuning starts. Here, the budget must change during the step, whenever inference reclaims memory.

Based on this observation, we present \system, a fine-grained memory sharing system that lets inference reclaim memory from a running tuning step one saved activation at a time. \system places the KV cache and the saved activations in one shared pool of physical GPU memory, and binds each saved activation to a fixed virtual-address range, called a \emph{slot}. A slot is \emph{backed} while physical pages of the pool are mapped behind it. When the KV cache needs more memory than the pool has free, \system unbacks slots and hands their pages to the KV cache while the tuning step continues. We call this mechanism \emph{\schemeB}.

Handing memory over inside a running step raises two challenges. First, the tuning step must remain correct although its slots can be unbacked at any point. With \emph{\schemeA}, \system checks whether a slot is backed before each kernel that uses it, saves activations only into backed slots, and recomputes every activation whose slot is unbacked. Second, a handover must never unback a slot that GPU work may still access. Because the CPU dispatches kernels ahead of the GPU~\cite{cudaasync}, a kernel dispatched several layers earlier may still access a slot when a handover arrives. With \emph{\schemeC}, \system protects every such slot, limits how far CPU dispatch runs ahead to keep the protected slots few, and makes every rank of a tensor-parallel deployment unback the same slots.

We evaluate \system on \rHeadEvalScope, where it keeps inference SLO attainment at or above \rHeadSloMolt while tuning continues. Colocation that shares only compute reaches an SLO attainment of at most \rHeadSloLms. Against the discard-based memory sharing of \sirius, which also lets inference reclaim memory, \system completes \rHeadTputGain the tuning work. On both \mistralsmall deployments, \moltstep, which moves the boundary only between steps, attains the SLO less often and completes less tuning work, which shows that the gain comes from the finer granularity rather than from the implementation.

In summary, this paper makes the following contributions:
{\flushList{\labelitemi}%
\begin{itemize}
    \item We identify the GPU memory that autoscaling leaves idle in LLM serving, and show that existing memory sharing cannot use it, because its unit, one training sample, is too coarse and each handover discards the running tuning step.
    \item We propose \system, a fine-grained memory sharing system whose \schemeB lets inference reclaim memory from a running tuning step one saved activation at a time.
    \item We design \schemeA, which keeps the tuning step correct, and \schemeC, which makes each handover safe under CPU--GPU asynchrony and tensor parallelism.
    \item We evaluate \system against compute sharing, discard-based memory sharing and \moltstep, and show that it sustains inference SLO attainment while completing more tuning work than discard-based memory sharing.
\end{itemize}}

%% file: contents/2-background.tex
\section{Background}
\label{sec:bg}

\subsection{Parameter-Efficient Fine-Tuning}
\label{sec:bg-peft}

A deployed model is adapted continually~\cite{llmstation,continualsurvey}, and PEFT is how that adaptation is done~\cite{adapters,prefixtuning,prompttuning,lora}. PEFT freezes the pretrained weights and trains a small set of added parameters, an \emph{adapter}, which is why it has become the standard method for adapting large models~\cite{peftsurvey,qlora,punica,slora}. Hugging Face carries \rMotiAdaptersTotal repositories tagged as PEFT adapters, created between \rMotiAdaptersSpan, and adapter publication has been rising since 2024~\cite{llmstation}. Llama models alone account for \rMotiAdaptersLlama of them~\cite{llama3}.

PEFT reads the pretrained weights but never writes them. LoRA, its representative method, replaces the update to a weight matrix \(W \in \mathbb{R}^{d \times k}\) with a low-rank product~\cite{lora},
\[
h(x) = Wx + \Delta W x = Wx + BAx,
\]
where \(B \in \mathbb{R}^{d \times r}\), \(A \in \mathbb{R}^{r \times k}\), and \(r \ll \min(d, k)\). Training updates only \(A\) and \(B\), about \(r(d+k)\) parameters instead of \(d \times k\), and the gradients and optimizer state shrink with them. A single copy of the model can therefore be shared between an inference process and a tuning process, rather than each holding its own~\cite{zero}.

\subsection{Idle GPU Resources in Autoscaled Serving}
\label{sec:bg-infer}

\begin{figure}[t]
  \centering
  \begin{subfigure}{\columnwidth}
    \centering
    \includegraphics[width=\columnwidth]{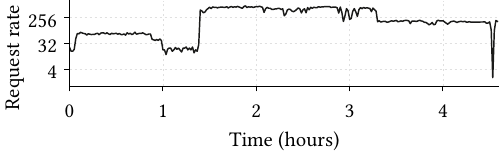}
    \caption{Request rate in the BurstGPT trace, requests per minute.}
    \label{fig:bg-traces-rate}
  \end{subfigure}
  \begin{subfigure}{\columnwidth}
    \centering
    \includegraphics[width=\columnwidth]{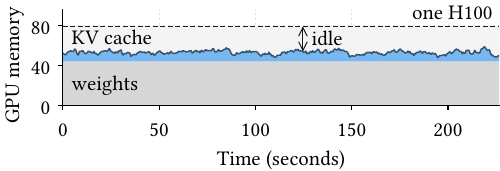}
    \caption{Memory on one \gpuh at TP=\rSetupTpSmall, in GiB.}
    \label{fig:bg-traces-tpone}
  \end{subfigure}
  \begin{subfigure}{\columnwidth}
    \centering
    \includegraphics[width=\columnwidth]{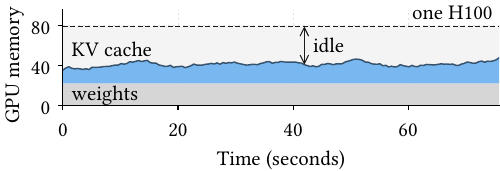}
    \caption{Memory on one \gpuh of two, at TP=\rSetupTpLarge, in GiB.}
    \label{fig:bg-traces-tptwo}
  \end{subfigure}
  \captionsetup{skip=\skipBgTraces}
  \caption{Request load over hours, and GPU memory over seconds.}
  \label{fig:bg-traces}
\end{figure}

A serving deployment follows request load by autoscaling, adding and removing GPUs to hold a utilization or concurrency target~\cite{vertexautoscale,gkeautoscale,kservekeda,rayserve}. Platforms set that target below full utilization, at \rMotiAutoscaleTarget GPU utilization by default on one of them~\cite{vertexautoscale}. The load moves slowly enough for this to work. Consecutive five-minute intervals of the BurstGPT trace~\cite{burstgpt} differ in request rate by a median of \rMotiRateIntervalSwing, and the rate steps between levels (Figure~\ref{fig:bg-traces}(a)). The variation that autoscaling absorbs is large. For instance, the load of user-facing services peaks during working hours and falls at night~\cite{burstgpt,dynamollm}. Over a week, peak load exceeds mean load by \rMotiDiurnalPeakMeanBurst in BurstGPT, and by \rMotiDiurnalPeakMeanConv and \rMotiDiurnalPeakMeanCode in the conversation and code services of the Azure trace~\cite{dynamollm}.

Autoscaling cannot follow the memory inside a replica, because that memory changes within seconds and adding capacity takes minutes. The scaling decision itself is prompt~\cite{gkeautoscale,kubehpa}. What takes minutes is bringing up a replica, whose GPUs hold the model split by tensor parallelism (TP)~\cite{megatron}, and loading the model into it~\cite{dynamollm,serverlessllm,sageserve}. When load falls, a replica that served a peak is kept for five more minutes by default~\cite{gkeautoscale,kubehpa}. In Figure~\ref{fig:bg-traces}(b), the KV cache of a \mistralsmall deployment covers \rMotiKvRangeGiB over \rMotiKvWindowSeconds and can grow \rMotiKvRiseTen within \rMotiKvSwingWindow.

Idle memory inside a replica cannot be returned to the provider, because scaling down removes a replica only as a whole, and its TP degree is fixed when it launches~\cite{loongserve}. A deployment runs the smallest number of replicas that meets the SLO at the peak of any rise in load that autoscaling cannot follow. At that autoscaling floor, capacity stands idle between peaks.  One \gpuh running \mistralsmall holds \rMotiWeightsGiB of weights and leaves \rMotiKvIdleGiB of its memory unused (Figure~\ref{fig:bg-traces}(b)). Under the heavier load of Figure~\ref{fig:bg-traces}(c), the KV cache needs \rMotiKvPeakTotalGiB against the \rMotiPoolTpOneGiB one GPU can offer beside the weights. The replica therefore spans two GPUs, which offer \rMotiPoolTpTwoGiB, and each of them holds \rMotiKvIdleGiBTp unused.

\subsection{Inference-first Tuning Colocation}
\label{sec:bg-coloc}

\begin{figure}[t]
  \centering
  \includegraphics[width=\columnwidth, trim=0 {\trimBurst} 0 0]{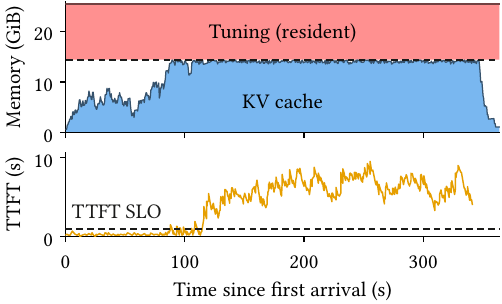}
  \caption{Memory and TTFT on a \mistralsmall deployment with tuning resident, under bursty arrivals.}
  \label{fig:limitation2}
\end{figure}

A second workload colocated with inference can use the memory idle at the autoscaling floor, but only until inference needs it again. When load rises, inference must reclaim that memory within seconds, before the arriving requests wait for it~\cite{vllm,orca}. This move across the boundary between the two workloads is a handover. The second workload must therefore tolerate interruption and hold a small \emph{resident footprint}, the memory that inference can never reclaim from it. PEFT meets both requirements (\S\ref{sec:bg-peft}). Its progress is measured in throughput rather than per-request latency, so it tolerates interruption. It shares the frozen weights inference already holds, so its resident footprint is small.

One line of colocation work shares compute. \llmstation~\cite{llmstation} decomposes tuning execution into fine-grained units and interleaves them with inference iterations, which protects serving latency while tuning runs. Sharing compute leaves the memory boundary fixed. As arrivals rise, the KV cache grows until it fills the memory that tuning leaves free, and TTFT exceeds the SLO once it does (Figure~\ref{fig:limitation2}).

The other line shares memory. \sirius~\cite{sirius} moves the boundary at runtime by resizing the training batch, removing samples when inference needs memory and adding them back once demand subsides. To make a handover fast enough for an inference SLO, it hands the freed pages straight to inference, bypassing the training framework's caching allocator~\cite{pytorch}. \S\ref{sec:moti} examines whether a boundary that moves at the granularity of whole training samples can follow the KV cache at the autoscaling floor.

%% file: contents/3-motivation.tex
\section{Motivation}
\label{sec:moti}

\subsection{Limitations of Existing Memory Sharing}
\label{sec:moti-limit}
\begin{figure}[!t]
  \centering
  \includegraphics[width=\columnwidth, trim=0 {\trimFootprint} 0 0]{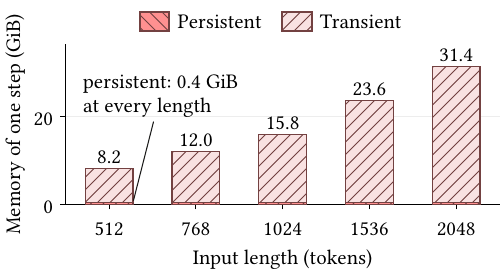}
  \caption{Memory of one batch-1 LoRA tuning step of \mistralsmall across input lengths, in live bytes.}
  \label{fig:limitation1}
\end{figure}

\sirius uses the idle memory that autoscaling removes, not the memory autoscaling leaves behind. Its target is the memory a fixed GPU count leaves free. This memory is large, so the boundary can move at the coarse granularity of whole training samples. It is also long-lived, so inference rarely takes it back and handovers are rare. Neither property holds at the autoscaling floor, where the idle memory is smaller and inference takes it back within seconds (\S\ref{sec:bg-infer}).

A first limitation is granularity. Within a batch, samples are independent, so adding or removing a whole sample cleanly adds or removes its memory. That memory is large relative to what the floor leaves idle. A \mistralsmall tuning step of one sample at \rMotiBatchOneFootprintSeq holds \rMotiBatchOneFootprintGiB beside the frozen base weights (Figure~\ref{fig:limitation1}), against the \rMotiKvIdleGiB left idle on one \gpuh at the floor (Figure~\ref{fig:bg-traces}(b)). There is no setting below one sample, so tuning halts entirely once the idle memory cannot hold one.

A second limitation is the cost of the handover itself. \sirius reclaims memory by discarding the current tuning step and restarting it with a smaller batch, so a handover late in a step wastes every operation the step has performed. At the floor, the KV cache can grow \rMotiKvRiseTen in \rMotiKvSwingWindow (Figure~\ref{fig:bg-traces}(b)), and each rise takes memory back from tuning. The cost of a discarded step is therefore paid at the rate the KV cache moves, not at the rate the day's load changes.

\subsection{Activation as the Unit of Memory Sharing}
\label{sec:moti-unit}
The limitations of \sirius point to two requirements for PEFT colocation. First, the sharing granularity must be much finer than a whole sample, so that idle GPU memory is used efficiently and tuning is not forced to halt the moment a single sample no longer fits. Second, a handover must preserve the progress of the running PEFT step rather than discard it, so that frequent handovers do not starve tuning.

The saved activation meets both. A saved activation is a tensor that the forward pass keeps for the backward pass to compute gradients. It is the only unit in a PEFT step that meets both requirements. Persistent state, namely the adapter parameters, gradients and optimizer state, must stay resident to preserve progress, and only \rMotiPersistentGiB of a step's memory is persistent, at every input length (Figure~\ref{fig:limitation1}). Intermediate tensors live too briefly for their memory to be reclaimed. Saved activations dominate the rest of the step's memory and can be recovered by recomputation~\cite{chen2016,checkmate,dtr,capuchin}. An activation is a fine-grained unit, because each one is small and can be dropped individually, at per-layer or per-operator granularity, rather than a whole sample at once. It also preserves progress, because a dropped activation is recomputed in the backward pass, and a handover then costs only that recomputation rather than the work that the step has already performed.

\subsection{Key Challenges of Fine-grained Memory Sharing}
\label{sec:moti-chal}
Using the saved activation as the sharing unit raises two challenges.

\textbf{C1. Adapting execution to a dynamically changing saved-activation set.} Because activations are dropped dynamically during the step, the backward pass cannot assume that a saved activation is still available. An activation may have been dropped at any point after it was produced, so reading it directly could lead to a memory fault. The system must therefore check, at each backward access, whether the activation is still in memory, and recompute it when it is not.

\textbf{C2. Dropping activations safely during execution.} The CPU decides which activations to drop, but it dispatches GPU kernels asynchronously and runs far ahead of the GPU execution~\cite{cudaasync,pytorch}, so a kernel it launched several layers back may still be reading or writing the activations of its layer. Freeing that memory would corrupt the kernel's work, and the CPU must protect the activations of every kernel it has launched but not seen complete~\cite{cudastreamalloc}, which shrinks the set it can drop. Under tensor parallelism, the CPU's lead over the GPU differs across ranks, so an activation can be safe to drop on one rank but not on another. Every rank must nevertheless drop the same activations, because tensor-parallel ranks must execute the same operators. The system must therefore make that decision at an arbitrary handover point. Dropping an activation that an in-flight kernel may still access causes a memory fault in the tuning step, and protecting one that no in-flight kernel can access withholds its memory from inference.

%% file: contents/4a-overview.tex
\section{\system}

\begin{figure}[t]
  \centering
  \includegraphics[page=2, width=\columnwidth]{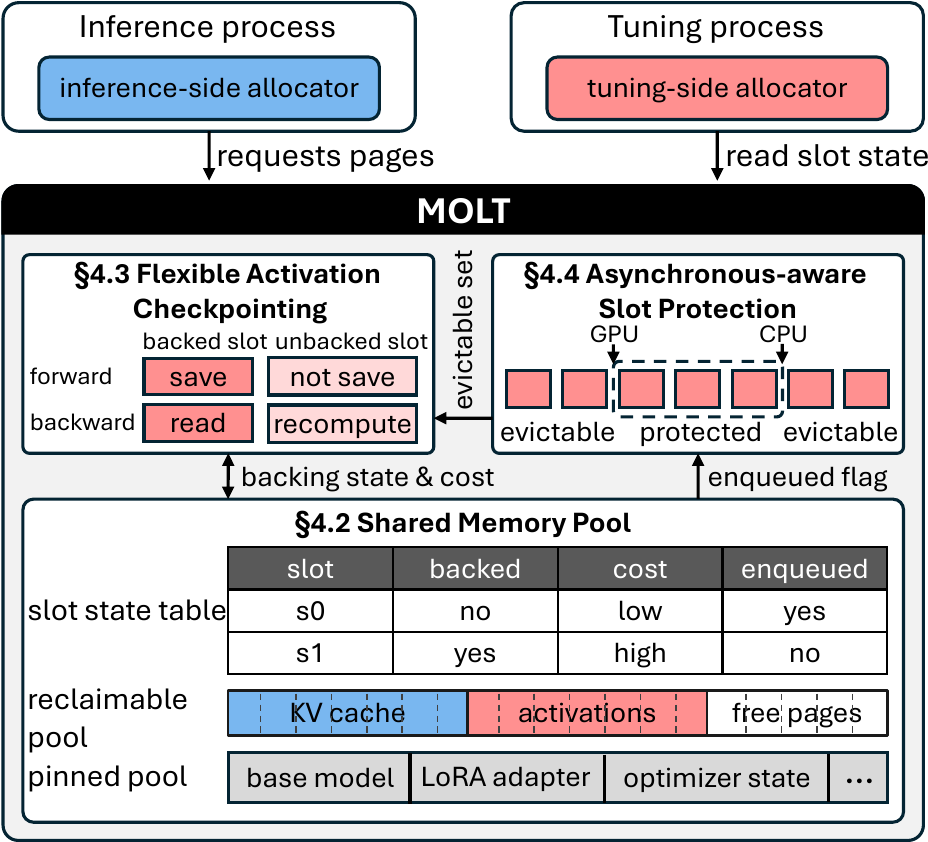}
  \caption{\system overview.}
  \label{fig:arch-overview}
\end{figure}

\subsection{Overview}
\label{sec:overview}
\system provides \schemeB, which allows inference to reclaim memory from a running tuning step at the granularity of a single saved activation. The tuning step continues across a handover, and its backward pass recomputes each activation whose memory was reclaimed, so no completed work is discarded.

\system is a fine-grained memory sharing system with three components (Figure~\ref{fig:arch-overview}). The \emph{inference-side allocator} and the \emph{tuning-side allocator} run inside vLLM and PyTorch, respectively (\S\ref{sec:eval-setup}). Each saved activation is bound to its slot, and the \emph{\sst} (\sstshort) records whether physical pages back each slot. The \emph{handover controller} manages the shared pool, which is preallocated and holds the KV cache and the pages of backed slots, the only memory a handover moves. When the KV cache exhausts the pool's free pages, the inference-side allocator requests pages over inter-process communication (IPC), and the handover controller unbacks slots and hands their pages to the KV cache.

With \schemeA, \system ensures that a tuning step remains correct when its slots are unbacked mid-step (C1, \S\ref{sec:moti-chal}). The tuning-side allocator reads a slot's backing state from the \sstshort before each kernel that produces or consumes the slot's activation (\S\ref{sec:flexible-ckpt}). The activation is saved in the forward pass only if its slot is backed, and in the backward pass, it is read from its slot if the slot is backed and recomputed otherwise. Unlike prior recomputation schemes~\cite{chen2016,checkmate,dtr}, \system decides which activations to recompute from backing state that changes whenever inference reclaims memory, while the step is still running.

With \schemeC, \system ensures that a handover unbacks no slot that in-flight GPU work may still access (C2, \S\ref{sec:moti-chal}). The \sstshort sets a slot's enqueued flag while a kernel that uses the slot has been dispatched but has not completed, and the handover controller unbacks only slots whose flag is clear on every rank (\S\ref{sec:async-slot-protection}).

%% file: contents/4b-design.tex
\begin{figure}[t]
  \centering
  \includegraphics[page=3, width=\columnwidth, trim=0 {\trimActLevelMem} 0 0]{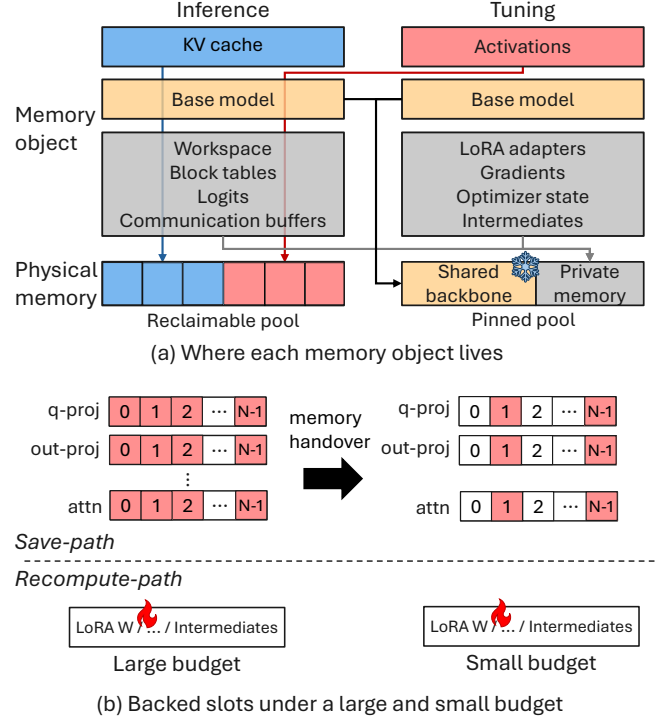}
  \caption{Reclaimable and pinned pools, with the tuning budget as the set of backed slots.}
  \label{fig:act-level-mem}
\end{figure}

\begin{figure*}[t]
  \centering
  \includegraphics[page=4, width=\textwidth, trim=0 {\trimHandoverMain} 0 0]{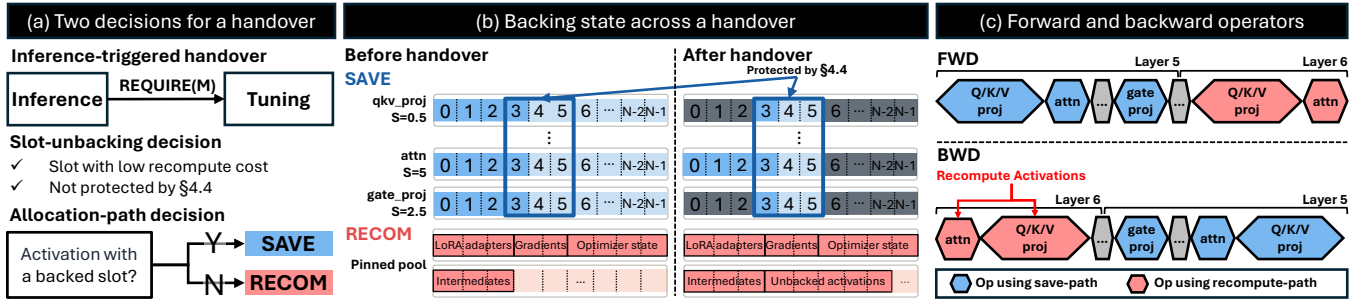}
  \caption{A handover in the middle of a step, which unbacks low-cost slots and switches only later operators to the \recomp.}
  \label{fig:cmh-main}
\end{figure*}

\subsection{Shared Memory Pool}
\label{sec:act-level-mem}

\system divides physical GPU memory into a \emph{\reclaimable}, the shared pool of \S\ref{sec:overview}, and a \emph{\pinned} for everything a handover does not move (Figure~\ref{fig:act-level-mem}(a)). The \pinned holds inference-side workspace and the frozen base-model weights, which inference and tuning read from the same physical pages. It also holds the LoRA parameters, gradients, optimizer state and intermediate tensors of tuning, and the activation of every unbacked slot, so each activation occupies exactly one pool. The \pinned holds an unbacked slot's activation only until its last use in the forward pass, and the backward pass recomputes it (\S\ref{sec:flexible-ckpt}).

\system represents the tuning budget as the set of backed slots (Figure~\ref{fig:act-level-mem}(b)). It reserves each slot once, before execution, from an offline profile of activation sizes, because the execution order of a tuning step is stable~\cite{swapadvisor,capuchin,vdnn}. Backing or unbacking a slot maps or unmaps the physical pages behind its fixed address range, so no activation is moved when the budget changes.

The \sstshort records three values for each slot (Figure~\ref{fig:arch-overview}). The handover controller writes the backing state when it maps or unmaps the slot's pages, and the tuning-side allocator reads it before each kernel that uses the slot (\S\ref{sec:flexible-ckpt}). The recomputation cost ranks the slots for unbacking and is fixed for each activation before execution (\S\ref{sec:flexible-ckpt}). \system never unbacks a slot whose activation cannot be recomputed, such as the input of a layer. The third value is the enqueued flag of \S\ref{sec:overview}, which keeps a handover from unbacking a slot that in-flight GPU work may still access (\S\ref{sec:async-slot-protection}).

\subsection{\schemeATitle}
\label{sec:flexible-ckpt}

\system lets a handover take effect at any point of a running tuning step and keeps the step correct (C1). A handover consists of two decisions, the \emph{slot-unbacking decision} of the handover controller and the \emph{allocation-path decision} of the tuning-side allocator (Figure~\ref{fig:cmh-main}(a)).

When inference requests $M$ bytes (\texttt{REQUIRE(M)} in Figure~\ref{fig:cmh-main}(a)), the handover controller takes the slot-unbacking decision and unbacks slots until they cover $M$ bytes. It considers only evictable slots, which \S\ref{sec:async-slot-protection} leaves unprotected because no in-flight GPU work can still access them. Among the evictable slots, it unbacks the slots with the lowest recomputation cost first, because every unbacked slot adds recomputation to the backward pass.

The tuning-side allocator takes the allocation-path decision when it dispatches each operator that produces a saved activation. If the activation's slot is backed, the activation follows the \emph{\save} (SAVE in the figure). The activation is allocated at the slot's address in the \reclaimable, where the backward pass reads it if the slot is still backed. If the slot is unbacked, the activation follows the \emph{\recomp} (RECOM in the figure). The activation is allocated in the \pinned, freed after its last use in the forward pass, and recomputed in the backward pass~\cite{chen2016,checkmate,dtr,korthikanti}. The allocator reads the backing state again when it dispatches each backward operator, so an activation whose slot is unbacked after it was saved is recomputed as well.

\begin{figure}[t]
  \centering
  \includegraphics[page=5, width=\columnwidth, trim=0 {\trimHandoverContrast} 0 0]{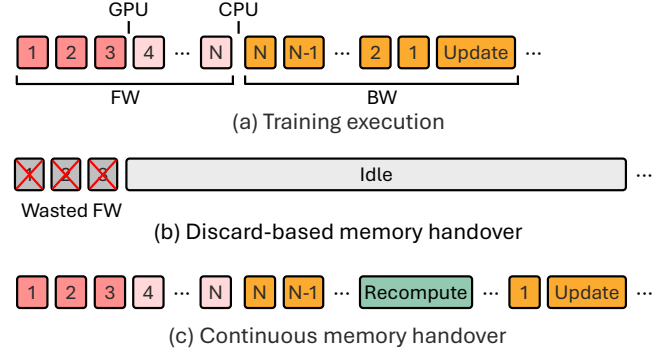}
  \caption{Forward work that discard-based handover wastes and \schemeB keeps, at the cost of recomputation.}
  \label{fig:cmh-contrast}
\end{figure}

In Figure~\ref{fig:cmh-main}(b), each row holds one kind of saved activation, each box is the slot of that activation in one layer, and $S$ is the recomputation cost of the row. Before the handover, every slot is backed. When the handover arrives, the CPU has dispatched the layers up to 5, but the GPU is still executing layer 3. In-flight GPU work may therefore still access the slots of layers 3 to 5, inside the box, so they are not evictable. The handover unbacks every other slot, so only these three layers keep backed slots. The activations that layers 0 to 2 have already saved are recomputed in the backward pass, and layers 6 onward allocate their activations in the \pinned. With a smaller handover request, the handover controller would stop earlier, and it would unback the slots of the \texttt{qkv\_proj} row first, because that row has the lowest $S$.

Every operator dispatched after the handover follows the new backing state (Figure~\ref{fig:cmh-main}(c)). The handover controller only writes the backing state, and the tuning-side allocator reads it when it dispatches an operator. Layers 3 to 5 were dispatched before the handover, so their slots remain backed, and the operators of layer 5 follow the \save, drawn in blue. Layer 6 is dispatched after the handover, when its slots are already unbacked, so its operators follow the \recomp, drawn in red. In the backward pass, layer 6 comes first, and \system recomputes its activations before its operators run. The operators of layer 5 then read their activations from their slots.

Unlike discard-based handover, \system never restarts the step, and a handover costs only the recomputation of the activations whose slots it unbacks (Figure~\ref{fig:cmh-contrast}). Discard-based handover couples every memory-boundary change to aborting and restarting the running step (\S\ref{sec:moti-limit}). In Figure~\ref{fig:cmh-contrast}(b), the forward work already done is discarded, and in Figure~\ref{fig:cmh-contrast}(c), it is kept and only the activations of unbacked slots are recomputed in the backward pass. Pages reach inference through the same handover path that discard-based systems use~\cite{sirius}.

Memory is handed back to tuning only at step boundaries, by \emph{re-backing} unbacked slots. At the start of each step, the handover controller maps free pages of the pool behind unbacked slots again, highest recomputation cost first, and stops when the pool has no free pages left. The remaining unbacked slots wait for a later step boundary. Re-backing a slot in the middle of a step would spare recomputation only for activations that the step has not produced yet. Waiting for the next step boundary therefore costs at most the extra recomputation of the rest of the current step. Because re-backing discards nothing, \system re-backs at every step boundary at which the pool has free pages. A discard-based system instead waits until free memory exceeds a threshold, because it moves the boundary only at a coarse granularity~\cite{sirius}.

\begin{figure}[t]
  \centering
  \includegraphics[page=6, width=\columnwidth, trim=0 {\trimAsyncThrottling} 0 0]{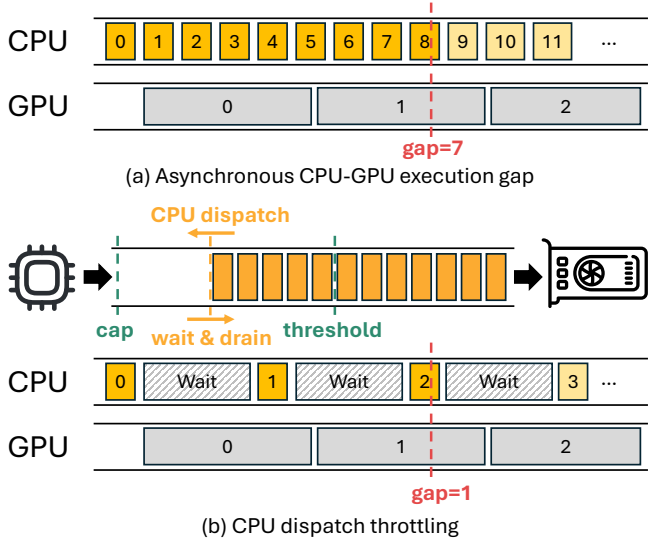}
  \caption{Dispatch throttling narrows the CPU--GPU gap, leaving fewer slots protected.}
  \label{fig:async-throttling}
\end{figure}

\subsection{\schemeCTitle}
\label{sec:async-slot-protection}
\begin{figure}[t]
  \centering
  \includegraphics[page=7, width=\columnwidth, trim=0 {\trimAsyncTp} 0 0]{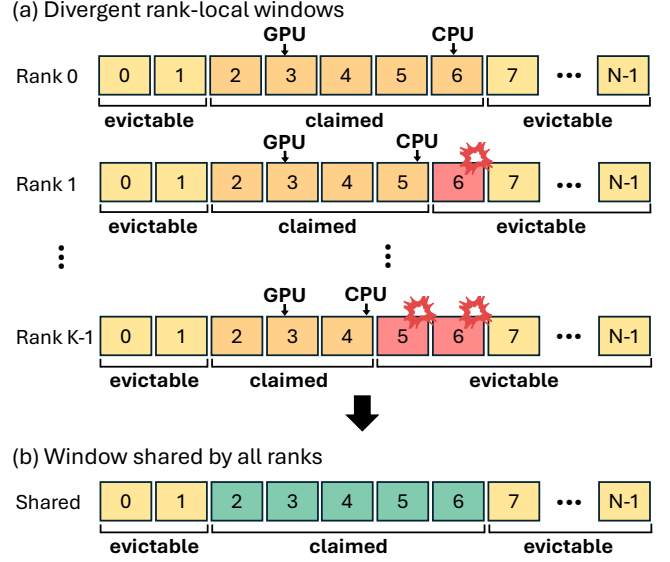}
  \caption{Every rank protects the union of the rank-local protection windows.}
  \label{fig:async-tp}
\end{figure}

\system protects every slot that in-flight GPU work may still read or write, and a handover unbacks only the other slots (C2). Under TP, every rank must also unback the same slots (Figure~\ref{fig:async-tp}). \system protects these slots without discarding the GPU work dispatched before the handover. \system classifies every slot by the enqueued flag that \S\ref{sec:overview} defines. A slot is \emph{claimed} while the enqueued flag of the slot is set. A handover leaves the backing state of a claimed slot unchanged, whether the slot is backed or was unbacked by an earlier handover. A slot is evictable when the enqueued flag of the slot is clear. 

The set of claimed slots, the \emph{protection window}, contains the slots of every layer from the layer before the \emph{completion frontier} to the \emph{dispatch frontier} (Figure~\ref{fig:async-throttling}(a)). The completion frontier is the layer that the GPU is executing, and the dispatch frontier is the last layer that the CPU has dispatched. The window includes the layer before the completion frontier because the first operators of a layer read the output of the previous layer. \system tracks GPU completion with a host-side callback, a CPU function that runs once the GPU has completed a layer~\cite{cudahostfunc}. \system sets the enqueued flag of every slot of a layer when the CPU dispatches the first operator of the layer. The enqueued flag covers a whole layer rather than one operator, because a later operator of the same layer, such as a residual connection, may read a slot that an earlier operator wrote~\cite{transformer}. \system clears the flag of a layer only when the host-side callback of the next layer runs. The dispatch frontier usually runs several layers ahead of the completion frontier, because CPU dispatch is asynchronous to GPU execution. In Figure~\ref{fig:async-throttling}(a), the CPU has dispatched layer 8 while the GPU still executes layer 1. The protection window thus contains layers 0 to 8, drawn in orange. The CPU has not yet dispatched layers 9 onward, whose slots are evictable and drawn in yellow.

When a handover request arrives, \system freezes the protection window until the handover controller has decided which slots to unback and has unbacked those slots. We call this mechanism the \emph{protection snapshot}. Without the protection snapshot, the CPU could dispatch an operator that uses one of the slots to unback, and that operator would then access a slot that has no physical pages. The protection snapshot therefore pauses CPU dispatch and holds the dispatch frontier in place. The pause has little effect on tuning, because the GPU continues to execute the operators already dispatched. The protection snapshot also defers the host-side callbacks that would advance the completion frontier, and thereby keeps every slot of the protection window claimed until the slots to unback are decided. A slot may remain claimed slightly longer than necessary, but no slot becomes evictable too early. After the handover controller has unbacked the chosen slots, \system runs the deferred callbacks and resumes CPU dispatch.

\system keeps the protection window small by limiting how far CPU dispatch may run ahead of GPU execution, which leaves more slots evictable when a handover request arrives (Figure~\ref{fig:async-throttling}(b)). We refer to this mechanism as \emph{dispatch throttling}. Without dispatch throttling, the protection window contains every layer that the CPU has dispatched ahead of the GPU, seven layers in Figure~\ref{fig:async-throttling}(a), and no slot in the protection window can be unbacked. \system enables dispatch throttling by modifying the software GPU kernel management technique used in \sirius~\cite{sirius}. The technique holds dispatched operators in a queue until the GPU executes the operators, and \system adds a \emph{dispatch cap} on the number of operators in the queue. When the queue reaches the dispatch cap, CPU dispatch waits until the GPU has executed enough operators to bring the queue below a threshold, which we set to half of the dispatch cap. In Figure~\ref{fig:async-throttling}(b), CPU dispatch waits after each layer, and the gap between the dispatch and completion frontiers shrinks from seven layers to one.

\system chooses the dispatch cap by offline profiling, as the smallest dispatch cap that does not lengthen the tuning step. A larger dispatch cap only widens the protection window, and a smaller dispatch cap leaves the GPU waiting for the CPU. The chosen dispatch cap therefore keeps the protection window as small as possible without slowing tuning.

Under TP, each rank keeps a separate protection window, and the handover controller unbacks only slots that lie outside the union of the rank-local protection windows (Figure~\ref{fig:async-tp}(b)). Collectives such as NCCL all-reduce~\cite{nccl} keep the GPUs of the ranks nearly aligned, but the dispatch frontiers of the ranks can differ (Figure~\ref{fig:async-tp}(a)). In the figure, every GPU executes layer 3, but rank 0 has dispatched layer 6, rank 1 layer 5, and rank $K-1$ layer 4. The slots of layer 6 are therefore claimed on rank 0 but evictable on ranks 1 and $K-1$, and the slots of layer 5 are evictable only on rank $K-1$, drawn in red. The collectives of the ranks would no longer match if every rank unbacked the slots that are evictable on that rank alone, because the ranks would then recompute different operators. The union of the rank-local protection windows, drawn in green, contains layers 2 to 6 on every rank. When a handover request arrives, every rank takes a protection snapshot and publishes the rank-local protection window through shared memory, and every rank then chooses the same slots to unback from the same union, without exchanging the chosen slots.

%% file: contents/5-evaluation.tex
\section{Evaluation}
\label{sec:evaluation}

We evaluate \system in inference-first colocation, where tuning uses the GPU memory that inference leaves idle and inference reclaims it whenever the KV cache needs more. The evaluation answers three questions.

{\flushList{3.}%
\begin{enumerate}
    \item Does inference keep its SLO attainment while \system completes more tuning work than discard-based memory sharing (\S\ref{sec:e2e-eval}, \S\ref{sec:e2e-mem})?
    \item Is the gain caused by handover granularity rather than by the implementation that carries it (\S\ref{sec:e2e-eval}, \S\ref{sec:granularity})?
    \item How much tuning speed does the dispatch cap cost, and how long does a handover delay inference (\S\ref{sec:dis-thr}, \S\ref{sec:ho-over})?
\end{enumerate}}

\begin{table}[t]
\centering
\setlength{\tabcolsep}{4pt}
\caption{Model and platform.}
\label{tab:model-deployments}
\begin{tabular}{lccc}
\toprule
Model & GPU & Layers & TP \\
\midrule
\mistralsmall  & \rSetupDeploySmall & \rSetupLayersSmall & \rSetupTpSmall \\
\mistralsmall  & \rSetupDeploySmallTptwo & \rSetupLayersSmall & \rSetupTpSmallTptwo \\
\llama      & \rSetupDeployLlamaSxm & \rSetupLayersLlama & \rSetupTpLlamaSxm \\
\llama      & \rSetupDeployLlama & \rSetupLayersLlama & \rSetupTpLlama \\
\bottomrule
\end{tabular}
\end{table}

\begin{figure}[t]
  \centering
  \includegraphics[width=\columnwidth, trim=0 {\trimOverview} 0 0]{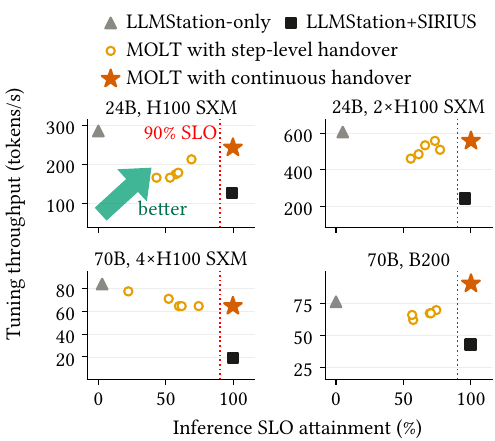}
  \caption{Overview of the end-to-end results.}
  \label{fig:eval-overview}
\end{figure}

\begin{table*}[t]
\centering
\caption{SLO attainment and tuning throughput under the BurstGPT trace.}
\label{tab:eval-e2e}
\begin{tabular}{@{}lrrrrrrrr@{}}
\toprule
 & SLO $\uparrow$ & Tuning $\uparrow$ & \multicolumn{3}{c}{TTFT (ms) $\downarrow$} & \multicolumn{3}{c}{TPOT (ms) $\downarrow$} \\
\cmidrule(lr){4-6}\cmidrule(l){7-9}
System & (\%) & (tokens/s) & p50 & p75 & p95 & p50 & p75 & p95 \\
\midrule
\multicolumn{9}{@{}l}{\textit{\mistralsmall{} (\rSetupDeploySmall)}} \\
\quad \llmstation-only & \rEteSxmLmsSlo & \rEteSxmLmsPeft & \rEteSxmLmsTtftMed & \rEteSxmLmsTtftPseventyfive & \rEteSxmLmsTtftPninetyfive & \rEteSxmLmsTpotMed & \rEteSxmLmsTpotPseventyfive & \rEteSxmLmsTpotPninetyfive \\
\quad \llmstation{}+\sirius & \rEteSxmSiriusSlo & \rEteSxmSiriusPeft & \rEteSxmSiriusTtftMed & \rEteSxmSiriusTtftPseventyfive & \rEteSxmSiriusTtftPninetyfive & \rEteSxmSiriusTpotMed & \rEteSxmSiriusTpotPseventyfive & \rEteSxmSiriusTpotPninetyfive \\
\quad \moltstep & \rEteSxmCoarseSlo & \rEteSxmCoarsePeft & \rEteSxmCoarseTtftMed & \rEteSxmCoarseTtftPseventyfive & \rEteSxmCoarseTtftPninetyfive & \rEteSxmCoarseTpotMed & \rEteSxmCoarseTpotPseventyfive & \rEteSxmCoarseTpotPninetyfive \\
\quad \textbf{\moltcont} & \rEteSxmMoltSlo & \rEteSxmMoltPeft & \rEteSxmMoltTtftMed & \rEteSxmMoltTtftPseventyfive & \rEteSxmMoltTtftPninetyfive & \rEteSxmMoltTpotMed & \rEteSxmMoltTpotPseventyfive & \rEteSxmMoltTpotPninetyfive \\
\midrule
\multicolumn{9}{@{}l}{\textit{\mistralsmall{} (\rSetupDeploySmallTptwo)}} \\
\quad \llmstation-only & \rEteSxmTptwoLmsSlo & \rEteSxmTptwoLmsPeft & \rEteSxmTptwoLmsTtftMed & \rEteSxmTptwoLmsTtftPseventyfive & \rEteSxmTptwoLmsTtftPninetyfive & \rEteSxmTptwoLmsTpotMed & \rEteSxmTptwoLmsTpotPseventyfive & \rEteSxmTptwoLmsTpotPninetyfive \\
\quad \llmstation{}+\sirius & \rEteSxmTptwoSiriusSlo & \rEteSxmTptwoSiriusPeft & \rEteSxmTptwoSiriusTtftMed & \rEteSxmTptwoSiriusTtftPseventyfive & \rEteSxmTptwoSiriusTtftPninetyfive & \rEteSxmTptwoSiriusTpotMed & \rEteSxmTptwoSiriusTpotPseventyfive & \rEteSxmTptwoSiriusTpotPninetyfive \\
\quad \moltstep & \rEteSxmTptwoCoarseSlo & \rEteSxmTptwoCoarsePeft & \rEteSxmTptwoCoarseTtftMed & \rEteSxmTptwoCoarseTtftPseventyfive & \rEteSxmTptwoCoarseTtftPninetyfive & \rEteSxmTptwoCoarseTpotMed & \rEteSxmTptwoCoarseTpotPseventyfive & \rEteSxmTptwoCoarseTpotPninetyfive \\
\quad \textbf{\moltcont} & \rEteSxmTptwoMoltSlo & \rEteSxmTptwoMoltPeft & \rEteSxmTptwoMoltTtftMed & \rEteSxmTptwoMoltTtftPseventyfive & \rEteSxmTptwoMoltTtftPninetyfive & \rEteSxmTptwoMoltTpotMed & \rEteSxmTptwoMoltTpotPseventyfive & \rEteSxmTptwoMoltTpotPninetyfive \\
\midrule
\multicolumn{9}{@{}l}{\textit{\llama{} (\rSetupDeployLlamaSxm)}} \\
\quad \llmstation-only & \rEteLlamaSxmLmsSlo & \rEteLlamaSxmLmsPeft & \rEteLlamaSxmLmsTtftMed & \rEteLlamaSxmLmsTtftPseventyfive & \rEteLlamaSxmLmsTtftPninetyfive & \rEteLlamaSxmLmsTpotMed & \rEteLlamaSxmLmsTpotPseventyfive & \rEteLlamaSxmLmsTpotPninetyfive \\
\quad \llmstation{}+\sirius & \rEteLlamaSxmSiriusSlo & \rEteLlamaSxmSiriusPeft & \rEteLlamaSxmSiriusTtftMed & \rEteLlamaSxmSiriusTtftPseventyfive & \rEteLlamaSxmSiriusTtftPninetyfive & \rEteLlamaSxmSiriusTpotMed & \rEteLlamaSxmSiriusTpotPseventyfive & \rEteLlamaSxmSiriusTpotPninetyfive \\
\quad \moltstep & \rEteLlamaSxmCoarseSlo & \rEteLlamaSxmCoarsePeft & \rEteLlamaSxmCoarseTtftMed & \rEteLlamaSxmCoarseTtftPseventyfive & \rEteLlamaSxmCoarseTtftPninetyfive & \rEteLlamaSxmCoarseTpotMed & \rEteLlamaSxmCoarseTpotPseventyfive & \rEteLlamaSxmCoarseTpotPninetyfive \\
\quad \textbf{\moltcont} & \rEteLlamaSxmMoltSlo & \rEteLlamaSxmMoltPeft & \rEteLlamaSxmMoltTtftMed & \rEteLlamaSxmMoltTtftPseventyfive & \rEteLlamaSxmMoltTtftPninetyfive & \rEteLlamaSxmMoltTpotMed & \rEteLlamaSxmMoltTpotPseventyfive & \rEteLlamaSxmMoltTpotPninetyfive \\
\midrule
\multicolumn{9}{@{}l}{\textit{\llama{} (\rSetupDeployLlama)}} \\
\quad \llmstation-only & \rEteLlamaBLmsSlo & \rEteLlamaBLmsPeft & \rEteLlamaBLmsTtftMed & \rEteLlamaBLmsTtftPseventyfive & \rEteLlamaBLmsTtftPninetyfive & \rEteLlamaBLmsTpotMed & \rEteLlamaBLmsTpotPseventyfive & \rEteLlamaBLmsTpotPninetyfive \\
\quad \llmstation{}+\sirius & \rEteLlamaBSiriusSlo & \rEteLlamaBSiriusPeft & \rEteLlamaBSiriusTtftMed & \rEteLlamaBSiriusTtftPseventyfive & \rEteLlamaBSiriusTtftPninetyfive & \rEteLlamaBSiriusTpotMed & \rEteLlamaBSiriusTpotPseventyfive & \rEteLlamaBSiriusTpotPninetyfive \\
\quad \moltstep & \rEteLlamaBCoarseSlo & \rEteLlamaBCoarsePeft & \rEteLlamaBCoarseTtftMed & \rEteLlamaBCoarseTtftPseventyfive & \rEteLlamaBCoarseTtftPninetyfive & \rEteLlamaBCoarseTpotMed & \rEteLlamaBCoarseTpotPseventyfive & \rEteLlamaBCoarseTpotPninetyfive \\
\quad \textbf{\moltcont} & \rEteLlamaBMoltSlo & \rEteLlamaBMoltPeft & \rEteLlamaBMoltTtftMed & \rEteLlamaBMoltTtftPseventyfive & \rEteLlamaBMoltTtftPninetyfive & \rEteLlamaBMoltTpotMed & \rEteLlamaBMoltTpotPseventyfive & \rEteLlamaBMoltTpotPninetyfive \\
\bottomrule
\end{tabular}
\end{table*}

\subsection{Experimental Setup}
\label{sec:eval-setup}

\textbf{Hardware.} We run \mistralsmall on \rSetupDeploySmall{} and \rSetupDeploySmallTptwo~\cite{h100}, and \llama on \rSetupDeployLlamaSxm{} and \rSetupDeployLlama~\cite{b200} (Table~\ref{tab:model-deployments}).

{\emergencystretch=3em \textbf{Models and adapters.} We tune rank-\rSetupLoraRank LoRA~\cite{lora} adapters on \mistralsmall~\cite{mistralsmall} and \llama~\cite{llama3}.\par}

\textbf{Datasets and traces.} We use ShareGPT~\cite{sharegpt} prompts for inference and the Alpaca~\cite{alpaca} dataset for tuning. A tuning sequence has \rSetupSeqSmall, \rSetupSeqSmallTptwo, \rSetupSeqLlamaSxm and \rSetupSeqLlama tokens in the deployments of Table~\ref{tab:model-deployments}, respectively. Inference requests arrive according to the BurstGPT trace~\cite{burstgpt}.

\textbf{Implementation.} \system is implemented as a patch to vLLM \rSetupVllm~\cite{vllm} and an extension library for \llmstation's build of PyTorch \rSetupTorch~\cite{pytorch}. The patch changes \rSetupVllmPatchLines lines in \rSetupVllmPatchFiles of vLLM's files and adds a \rSetupVllmModule-line module. The module ports \llmstation's execution interleaving~\cite{llmstation} to vLLM's V1 engine~\cite{vllmv1} and integrates \system with it. The \rSetupLocLib-line library backs and unbacks slots through CUDA virtual memory management~\cite{cudavmm} on \sirius's page pool~\cite{sirius}.

{\emergencystretch=3em \textbf{Baselines and ablation.} We compare \moltcont, the full design, against two baselines and one ablation. The ablation, \moltstep, differs from \moltcont only in when the boundary may move, so it isolates handover granularity. \llmstation-only shares only compute. It keeps tuning memory resident by excluding the same amount of memory from KV cache admission. \llmstation{}+\sirius also shares memory. At each handover, it resizes the training batch and discards the running step~\cite{sirius}. Its shared pool is the same size as that of \system. At each step boundary, \moltstep reserves memory for inference. The reserve is the KV cache that the waiting requests need, multiplied by a reserve factor, plus a fixed margin. We report the factor with the highest SLO attainment among \rAblFactorCount factors from \rSetupFactorMin to \rSetupFactorMax (\S\ref{sec:granularity}).\par}

\textbf{Metrics and SLOs.} We report SLO attainment and tuning throughput over a measurement window. SLO attainment is the fraction of requests arriving in the window that meet the SLO~\cite{alpaserve,distserve}. Tuning throughput is the number of tokens in the tuning samples completed in the window, per second. A request meets the SLO when its TTFT and its time per output token (TPOT) are each within \rSetupSloMultStrict the mean for the same deployment running inference alone~\cite{nestedfp,sola,sarathi,splitwise}.

\subsection{End-to-end Evaluation}
\label{sec:e2e-eval}

{\emergencystretch=3em \moltcont is the only configuration whose SLO attainment and tuning throughput are both high on every deployment. In Figure~\ref{fig:eval-overview}, \moltcont alone lies in the upper right on every deployment. The TTFT and TPOT percentiles in Table~\ref{tab:eval-e2e} are over the requests in the measurement window.\par}

\begin{figure*}[t]
  \centering
  \includegraphics[width=0.95\textwidth, trim=0 {\trimMemutil} 0 0]{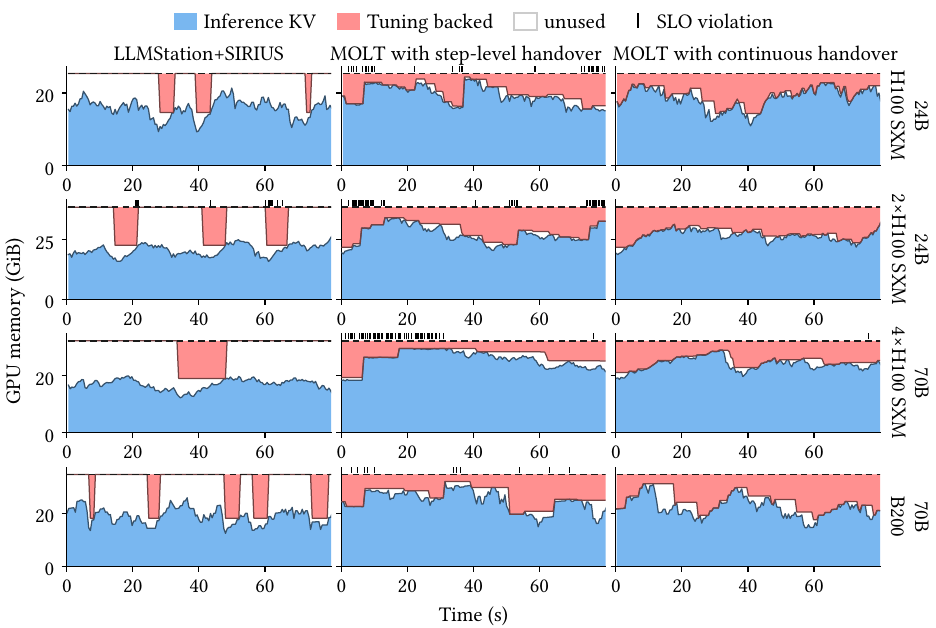}
  \caption{Shared-pool usage under the BurstGPT trace, where \llmstation{}+\sirius leaves memory unused that \moltcont gives to tuning.}
  \label{fig:eval-memutil}
\end{figure*}

\textbf{SLO attainment.} \moltcont meets the SLO for at least \rHeadSloMolt of requests on every deployment, at the right edge of Figure~\ref{fig:eval-overview}, and \llmstation-only for at most \rHeadSloLms, at the left edge. \llmstation{}+\sirius also meets it for at least \rEteSxmTptwoSiriusSlo\%, because it also moves memory to inference at each handover. \llmstation-only keeps tuning memory resident, so requests wait for the memory that tuning holds (\S\ref{sec:bg-coloc}). The TTFT p50 of \llmstation-only is at least \rEteSxmTptwoLmsTtftMed\,ms on every deployment, against at most \rEteLlamaBMoltTtftMed\,ms under \moltcont (Table~\ref{tab:eval-e2e}).

\textbf{Tuning throughput.} Of the configurations that meet the SLO, \moltcont completes the most tuning work. It completes \rEteMoltOverSiriusRange the tuning work of \llmstation{}+\sirius on every deployment (Figure~\ref{fig:eval-overview}). The ratio is smallest on \mistralsmall at TP=\rSetupTpSmall, \rEteSxmMoltPeft against \rEteSxmSiriusPeft tokens/s, and largest on \llama at TP=\rSetupTpLlamaSxm, \rEteLlamaSxmMoltPeft against \rEteLlamaSxmSiriusPeft tokens/s (Table~\ref{tab:eval-e2e}). At each handover, \llmstation{}+\sirius discards the running step, and \moltcont preserves it. \S\ref{sec:e2e-mem} counts the discarded steps. \llmstation-only tunes faster on both \mistralsmall deployments, at \rEteSxmLmsPeft and \rEteSxmTptwoLmsPeft tokens/s, because its tuning memory is never reclaimed.

\textbf{Source of the gain.} Handover granularity alone explains why \moltcont outperforms \moltstep on both \mistralsmall deployments. The two differ only in when the boundary may move (\S\ref{sec:eval-setup}). On \mistralsmall at TP=\rSetupTpSmall, \moltcont meets the SLO for \rEteSxmMoltSlo\% of requests at \rEteSxmMoltPeft tokens/s, against \rEteSxmCoarseSlo\% at \rEteSxmCoarsePeft tokens/s for \moltstep. At TP=\rSetupTpSmallTptwo, \moltcont also attains more and tunes faster. Its TTFT p95 at TP=\rSetupTpSmall is \rEteSxmMoltTtftPninetyfive\,ms against \rEteSxmCoarseTtftPninetyfive\,ms, and \S\ref{sec:e2e-mem} shows where the requests of \moltstep wait. On \llama at TP=\rSetupTpLlamaSxm, the gap is in SLO attainment only. \moltcont attains \rEteLlamaSxmMoltSlo\% against \rEteLlamaSxmCoarseSlo\% at the same \rEteLlamaSxmCoarsePeft tokens/s. 

\subsection{Memory Utilization}
\label{sec:e2e-mem}

\begin{figure*}[t]
  \centering
  \includegraphics[width=\textwidth, trim=0 {\trimCapsweep} 0 0]{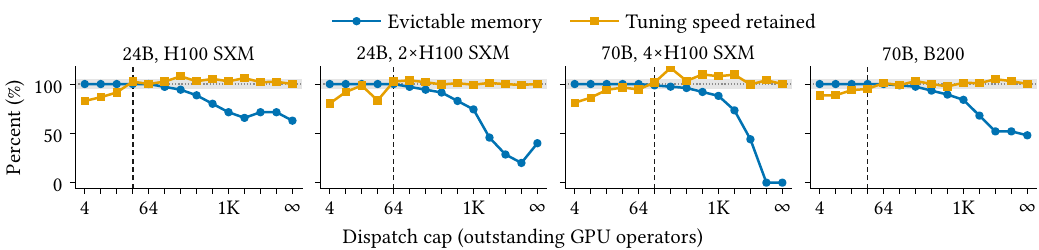}
  \caption{Evictable memory and tuning speed retained against the dispatch cap, both high at the dashed cap.}
  \label{fig:eval-cap-sweep}

\end{figure*}

\begin{figure*}[t]
  \centering
  \includegraphics[width=\textwidth, trim=0 {\trimHoover} 0 0]{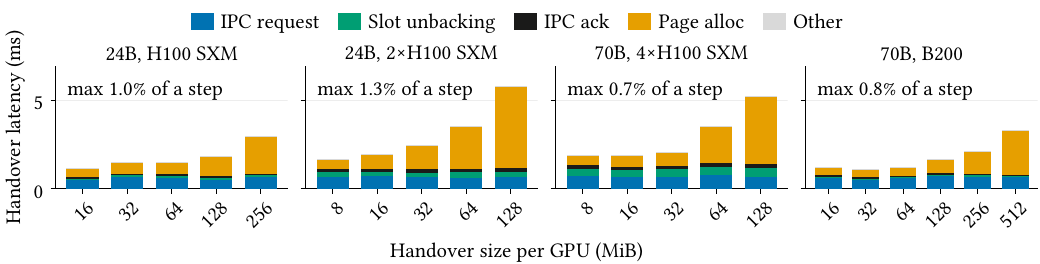}
  \caption{Mean handover latency by stage against handover size, growing with inference-side page allocation.}
  \label{fig:eval-ho-overhead}
  
\end{figure*}

At the memory boundary, \moltcont gives tuning most of the memory that inference leaves free, and inference rarely waits for memory. The end-to-end differences of \S\ref{sec:e2e-eval} arise there. Figure~\ref{fig:eval-memutil} draws how inference and tuning divide the shared pool over the first \rMemPlotSpan of each deployment's measurement window, on one time axis. The inference KV cache is drawn up from zero. The tuning-backed memory is drawn down from the pool size, which is dashed. A tick above the pool marks the arrival of each request that misses the SLO. Every configuration of a deployment ran at one pool size, so a row compares them on one memory axis. The percentages below are over the whole measurement window, and the attainable budget is the pool minus the inference KV cache.

\textbf{Memory use.} \moltcont uses most of the attainable budget for tuning, and \llmstation{}+\sirius leaves most of it unused, in every row of Figure~\ref{fig:eval-memutil}. On \mistralsmall at TP=\rSetupTpSmall, \moltcont backs \rMemMoltBackedMeanPct of the attainable budget on average, and \moltstep backs \rMemCoarseBackedMeanPct. In contrast, \llmstation{}+\sirius leaves \rMemSiriusUnusedPct of that budget unused. \moltcont unbacks slots within a step, and it re-backs them only at a later step boundary, because waiting costs at most the recomputation of the rest of the step (\S\ref{sec:flexible-ckpt}). In Figure~\ref{fig:eval-memutil}, its tuning-backed memory therefore shrinks within a step as the KV cache grows, and grows back only at step boundaries. The tuning-backed memory of \moltstep, however, changes only at step boundaries, in both directions. \llmstation{}+\sirius gives tuning a discarded step's memory again only after free memory exceeds a threshold (\S\ref{sec:flexible-ckpt}), so it gives tuning either all of that memory or none of it. It discards \rMemWastedFracSirius of its step attempts.

\textbf{Waiting for memory.} \moltcont rarely makes inference wait for memory, and \moltstep often does. We sample the inference engine's request queue every \rSetupPollInterval. On every deployment, a request was waiting for memory in at most \rMemQueuedPctMoltMax of the samples under \moltcont, against \rMemQueuedPctCoarseRange under \moltstep. Almost all SLO misses of \moltstep exceed the TTFT limit, so its ticks in Figure~\ref{fig:eval-memutil} mark requests that waited for their first token. \moltcont and \llmstation{}+\sirius both give inference memory within a step, and the handovers of \moltcont give inference the requested memory within \rMemMoltTrackLatency at p99. \moltstep gives inference memory only at step boundaries. A request whose KV cache does not fit in free memory therefore waits until the running step ends.

\subsection{Granularity of the Memory Boundary}
\label{sec:granularity}

\moltcont meets the SLO for more requests than \moltstep at every reserve factor on every deployment (Figure~\ref{fig:eval-overview}). Only on \llama at TP=\rSetupTpLlamaSxm does a factor tune faster than \moltcont. The factor sets how much memory \moltstep reserves for inference at each step boundary (\S\ref{sec:eval-setup}). On \mistralsmall at TP=\rSetupTpSmall, the middle factor, \rAblBestFactor, is the best on both axes, and Table~\ref{tab:eval-e2e} reports it. The other four attain as little as \rAblWorstSlo\%, and neither axis is monotonic in the factor.

\moltcont moves the boundary far more often than \moltstep, and no reserve factor removes the waits for memory that follow. On \mistralsmall at TP=\rSetupTpSmall, \moltcont handed memory to inference \rAblHandoversFine times over each whole run, against \rAblHandoversCoarse for \moltstep at factor \rAblBestFactor. Even at factor \rAblBestFactor, a request was waiting for memory in \rMemQueuedPctCoarse of the samples (\S\ref{sec:e2e-mem}). At the other four factors, this share is higher, at \rMemQueuedPctCoarseOthers.

\subsection{Dispatch Throttling Analysis}
\label{sec:dis-thr}

A dispatch cap can leave most saved activations evictable without slowing tuning, on every deployment. The cap trades evictable memory against tuning speed (\S\ref{sec:async-slot-protection}), and Figure~\ref{fig:eval-cap-sweep} sweeps it on each deployment of Table~\ref{tab:model-deployments}. Each sweep runs tuning without inference at the deployment's tuning sequence length. Evictable memory is the share of layers outside the protection window at its widest. Its base excludes the \rCapBackwardFloor layers that the backward pass keeps claimed at every cap, so it is 100\% when the window is at its minimum. The point at $\infty$ is the unthrottled run, without a cap. Tuning speed retained is the tuning speed at the cap as a share of the unthrottled speed. The shaded band is $\pm$\rCapTolerancePct of the unthrottled speed, a tolerance we set rather than a spread we measured. Each point is one run. The dashed line marks the smallest cap from which no larger cap falls below the band.

At the dashed cap, tuning speed stays within the band, and the protection window at its widest spans \rCapClaimedAtBandPct of all layers, close to its minimum, against \rCapClaimedNoCapPct without a cap, in the order of Table~\ref{tab:model-deployments}. The dashed cap is \rCapBandCaps. Tuning slows at the smallest caps. At a cap of \rCapTightest, a step takes \rCapStepTightRange longer than unthrottled.

\subsection{Handover Overhead Analysis}
\label{sec:ho-over}

A handover takes at most \rHeadHoOverhead of a tuning step, and only page allocation on the inference side grows with the handover size. A handover delays the inference request that needs its pages, from the page request of the inference-side allocator to the completion of page allocation. Under \moltstep, that request waits instead until the running step ends (\S\ref{sec:e2e-mem}), so we state the latency as a share of a tuning step. In Figure~\ref{fig:eval-ho-overhead}, each bar is the mean latency of at least \rHoRepsMin handovers of one size, broken into four stages and Other, the time between them. In time order, the inference-side allocator requests pages over IPC (IPC request), and the handover controller takes the slot-unbacking decision and unbacks the chosen slots (Slot unbacking). The controller then acknowledges over IPC (IPC ack), and the allocator allocates their pages to the KV cache (Page alloc). At the largest size drawn for each deployment, \rHoSizeDrawnList in the order of Table~\ref{tab:model-deployments}, the mean handover takes \rHoLargestShareList of a step.

Three stages change little with the handover size: the IPC request, slot unbacking and the IPC acknowledgment. They take \rHoFixedTpOne on the two deployments at TP=\rSetupTpSmall and \rHoFixedTpMulti on the two with tensor parallelism, and they are more than half of a handover at the smallest size on every deployment. Of the three, slot unbacking grows the most with the tensor-parallel degree, yet it remains within \rHoPlanStepPct of a tuning step at every degree. It takes \rHoPlanTpOne at TP=\rSetupTpSmall, \rHoPlanTpTwo at TP=\rSetupTpSmallTptwo and \rHoWorkLlamaSxm at TP=\rSetupTpLlamaSxm. It grows because, under tensor parallelism, the slot-unbacking decision also waits for every rank's protection snapshot (\S\ref{sec:async-slot-protection}).

Page allocation changes little at small handover sizes and grows almost in proportion to the size at large ones. In Figure~\ref{fig:eval-ho-overhead}, on both deployments at TP=\rSetupTpSmall, allocation remains within \rHoAllocFlatTpOne up to \rHoAllocKneeSize per GPU. Beyond that size, each doubling of the size multiplies it by \rHoAllocDoublingTpOne. The growth comes from allocation taking each page from the free list of the shared pool and zero-filling it. At TP=\rSetupTpSmallTptwo, allocation grows from the smallest size drawn. At TP=\rSetupTpLlamaSxm, it rises at \rHoPaceSizeLlamaSxm, the first size measured with a longer interval between handovers.

%% file: contents/6-related.tex
\section{Related Work}
\label{sec:related}

\textbf{Colocation of LLM inference and tuning.} Unlike prior colocation systems, \system reclaims memory from a running tuning step without discarding the step. FineInfer runs inference between tuning steps~\cite{fineinfer}. FlexLLM interleaves tuning and inference tokens in each iteration and reserves the KV cache statically~\cite{flexllm}. \llmstation interleaves tuning operations with inference iterations while tuning memory remains resident (\S\ref{sec:bg-coloc}). \sirius moves the memory boundary by discarding the running step and restarting it with a smaller batch (\S\ref{sec:moti-limit}). \system reuses \llmstation's interleaving and \sirius's page pool (\S\ref{sec:eval-setup}), but its unit of memory sharing is a saved activation rather than a training sample.

\textbf{Activation checkpointing.} \system builds on activation checkpointing and differs from it in the memory budget, which inference changes while the tuning step runs (\S\ref{sec:overview}). Checkpointing chooses which activations to recompute under a budget fixed before training starts~\cite{chen2016,checkmate,korthikanti,gruslys}. DTR and Capuchin decide at run time under such a budget~\cite{dtr,capuchin}. Like DTR~\cite{dtr} and selective recomputation~\cite{korthikanti}, \system gives up the activations that are cheapest to recompute first (\S\ref{sec:flexible-ckpt}). In these systems, the memory an activation gives up stays with the training job. In \system, inference reclaims it while dispatched kernels may still access it, which \schemeC makes safe (\S\ref{sec:async-slot-protection}).

%% file: contents/7-con.tex
\section{Conclusion}
\label{sec:conclusion}

Autoscaling leaves GPU memory idle inside LLM serving replicas, because it cannot follow a KV cache that changes within seconds and removes a replica only as a whole. Existing colocation systems either keep the tuning memory resident or let inference reclaim it only in whole training samples, discarding the running tuning step each time. We present \system, a fine-grained memory sharing system for colocating LLM inference and tuning. \system lets inference reclaim memory one saved activation at a time, safely under CPU--GPU asynchrony and tensor parallelism, while the running step continues and recomputes those activations. Experiments show that \system keeps inference SLO attainment at or above \rHeadSloMolt while completing \rHeadTputGain the tuning work of discard-based memory sharing.

\section*{Acknowledgments}

This work used Claude Code for code, language editing and text refinement. All technical content, experiments, and conclusions were verified by the authors.